%% file: paper.tex
\documentclass{onethree}

\definecolor{headingcolor}{HTML}{000000} 
\renewcommand{\onethreeblue}[1]{{%
  \bfseries\color{headingcolor}#1%
}}

\renewcommand{\footnoterule}{%
  \kern -3pt
  \hrule width 0.8\columnwidth height 0.4pt
  \kern 2.6pt
}
\hypersetup{
  linkcolor=black, 
  urlcolor=black    
}
\usepackage[toc,page,header]{appendix}
\usepackage{minitoc}
\usepackage{amsmath,amssymb}
\usepackage{bm}
\usepackage{wrapfig}
\usepackage{enumitem}
\usepackage[table]{xcolor}
\usepackage[most]{tcolorbox}
\usepackage{xcolor}
\usepackage{booktabs}
\usepackage{graphicx}
\usepackage[table]{xcolor}
\usepackage[most]{tcolorbox}
\usepackage{xcolor}
\usepackage{booktabs}
\usepackage{multirow} 
\usepackage{listings}
\usepackage{cuted}
\usepackage{stfloats}
\usepackage{slashed}
\usepackage{pifont}
\usepackage{enumitem}

\usepackage{times}

\renewcommand{\titlefont}{\fontsize{17}{20}\selectfont\bfseries}
\DeclareTextFontCommand{\textbf}{\bfseries}

\newtcolorbox{examplebox}{
  colback=gray!5,
  colframe=gray!55,
  boxrule=0.8pt,
  arc=4mm,
  left=5mm,
  right=5mm,
  top=3mm,
  bottom=3mm,
  breakable
}

\title{LLMs Can Tackle High Energy Theoretical Physics, yet Gaps Remain: Part I, Quantum Field Theory}

\author[1,6]{Tong Liu$^{*}$}
\author[2]{Bohan Li$^{*}$}
\author[3]{Gongrui Cheng$^{*}$}
\author[4]{Zixiong Yu}
\author[5]{Yulan He}
\author[1,6]{Volker Tresp}

\renewcommand{\affiliationformat}[2][]{{%
  \affiliationfont
  \fontfamily{ptm}\fontseries{m}\fontshape{n}\selectfont
  \textsuperscript{#1}#2%
}}

\renewcommand{\affiliationlist}{%
  \affiliationformat[1]{\textit{Ludwig-Maximilians-Universität München, München, Germany}}\\
  \affiliationformat[2]{\textit{Okinawa Institute of Science and Technology Graduate University, 
Okinawa, Japan}}\\
  \affiliationformat[3]{\textit{Instituut voor Theoretische Fysica, K.U. Leuven, 
  Leuven, Belgium}}\\
  \affiliationformat[4]{\textit{Tsinghua University}}
  \affiliationformat[5]{\textit{King’s College London}}
  \affiliationformat[6]{\textit{Munich Center for Machine Learning}}
}

\abstract{
\vspace{-0.15cm}
Beyond math and code, Large language models (LLMs) are beginning to tackle problems that have traditionally required substantial training for theoretical physicists, but it remains unclear how far this capability extends and where it might break. 
As a first step toward pushing LLMs in high-energy theoretical physics, we introduce QFT-Bench, a benchmark of 75 graduate- to research-level problems requiring extensive reasoning and derivation across core Quantum Field Theory (QFT) subareas.
Each question has a well-defined target and closed-form reference answer. 
Evaluation combines symbolic verification with LLM-based judge. 
Across 39 proprietary and open-source models, frontier models already demonstrate substantial capabilities on nontrivial QFT calculations, with the strongest models exceeding 95\% accuracy. 
At the same time, clear gaps remain. 
Even strong models still make errors in calculations, conventions, and instruction following, while weaker models often fail at the level of physical concepts or derivations. 
We further monitor models' CoTs and show that current models exhibit nontrivial reasoning unfaithfulness, i.e., they could reach correct answers but the reasoning itself does not justify and logically support them. 
These results provide a quantitative assessment of current LLM capabilities in QFT and a first step toward broader evaluation in high-energy theoretical physics.
\vspace{-0.1cm}
}

\checkdata[Resources]{
\href{https://huggingface.co/datasets/LLMsHEP/QFT-Bench}{\raisebox{-1ex}{\includegraphics[height=3.3ex]{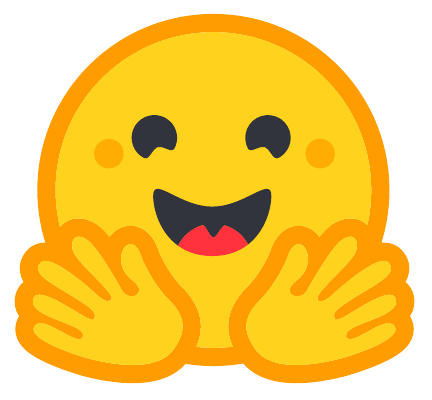}}~Data}
\hspace{1em}
\href{https://qft-bench.github.io/}{\raisebox{-1ex}{\includegraphics[height=3.2ex]{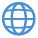}}~Page}
}

\titleformat*{\section}
  {\normalfont\fontsize{12}{14}\selectfont\bfseries}

\titleformat*{\subsection}
  {\normalfont\fontsize{10}{11}\selectfont\bfseries}

\titleformat*{\subsubsection}
  {\normalfont\fontsize{10}{11}\selectfont\bfseries}

\setcitestyle{authoryear,round,semicolon}

\begin{document}
\vspace{-2cm}
\maketitle
\vspace{-1.1cm}

\begingroup
\renewcommand{\thefootnote}{*}
\footnotetext{Equal contribution. E-mail: tongliu.physics@gmail.com\, bohan.li@oist.jp\, gongrui.cheng@kuleuven.be}
\endgroup

\begin{figure}[!htbp]
\vspace{-0.1cm}
\hspace{-0.4cm}
\centering
\includegraphics[width=0.98\linewidth]{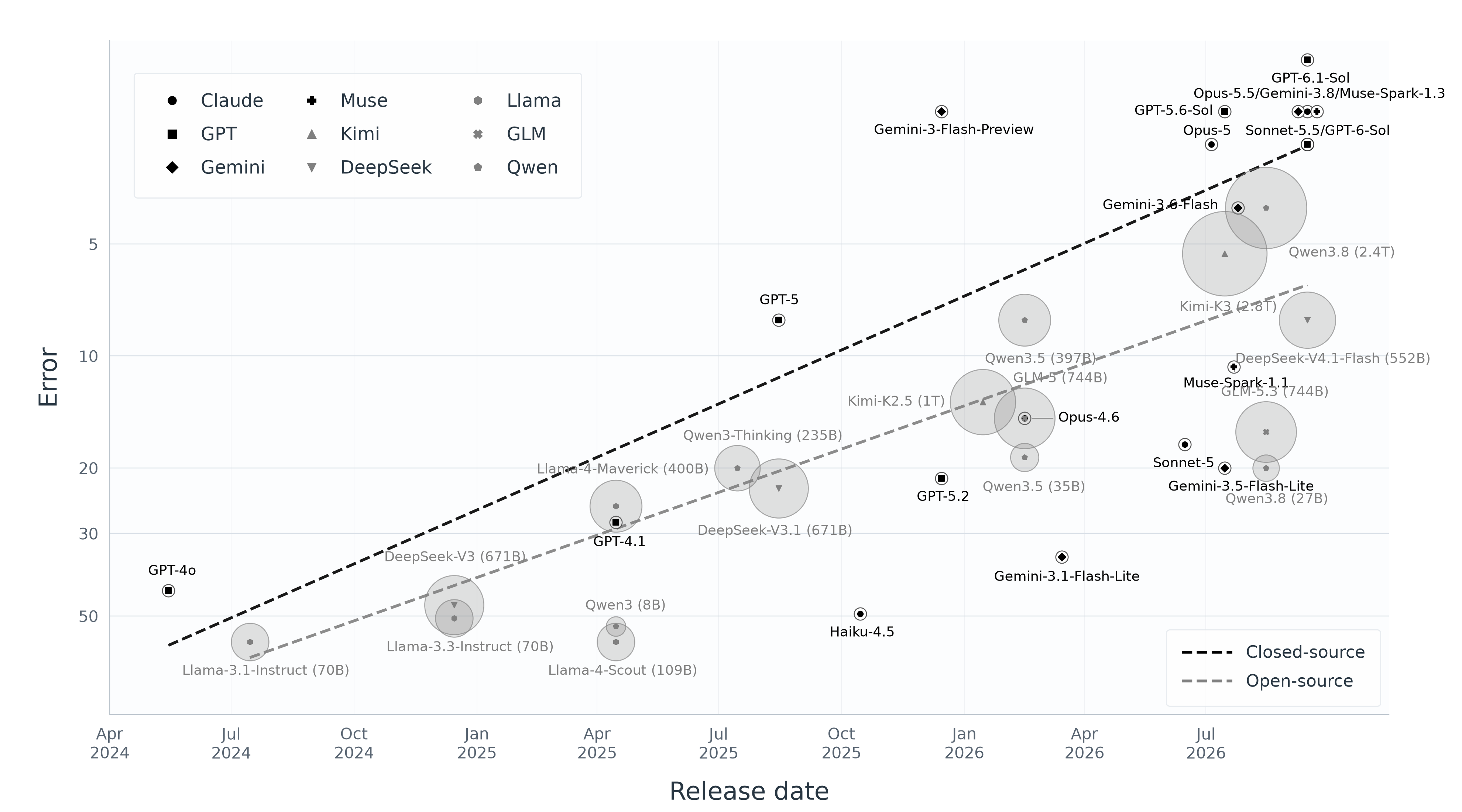}
\hspace{-0cm}
\vspace{-0.3cm}
\caption{\textbf{LLM Scaling Law on QFT-Bench.} }
\label{fig:teaser}
\end{figure}

\newpage

\rmfamily
\fontsize{11pt}{13.5pt}\selectfont

\tableofcontents
\newpage

\input{sections/introduction}

\input{sections/relatedwork}

\input{sections/data}

\input{sections/experiments}

\input{sections/results}

\input{sections/conclusion}

\clearpage

\bibliographystyle{plainnat}
\bibliography{main}

\clearpage

\beginappendix
\input{sections/appendix}

\end{document}

%% file: sections/introduction.tex
\section{Introduction}

Large language models (LLMs) have advanced rapidly in mathematical reasoning: moving from solving Olympiad-level problems to making substantive contributions to open mathematical questions~\citep{openai_ten, alpoge2026more,openai_ns,openai_sharing} within a remarkably short period. 
One plausible reason behind this progress is that mathematics, as well as domains such as code, \textit{often comes with cheap, mechanical and binary verification}, providing unusually reliable signals for model training. 
For example, formal systems such as Lean~\citep{de2015lean} break down proofs into smaller and verifiable components, enabling models to learn through iterative exploration and refinement. 
This helps explain why LLMs remain substantially less capable on discovery-oriented natural science such as theoretical physics, where 
\textit{(1). problems are less formally specified; 
(2). breakthroughs require fundamentally new ways of understanding nature; 
(3). the validity of a theoretical result may depend on physical plausibility, consistency with established theory, and ultimately the test of experimental evidence.}

Quantum field theory (QFT) makes an especially demanding test case within this gap. 
It is the mathematical language of the Standard Model and the core of much of modern theoretical physics~\citep{peskin1995introduction,weinberg1995quantum}. 
The reasoning itself is hard: QFT calculations often involve long, multi-step derivations in which a small error can change the result, and they require more than correct algebra, since one must identify the physical assumptions and approximations that apply. 
Moreover, QFT is underrepresented in training data: compared with mathematics and code, it offers much fewer verifiable training samples, and the available text is highly technical and fragmented across subfields. 

We introduce QFT-Bench, a set of 75 hard-graduate- to easy-research-level quantum field theory problems spanning eight subareas, from classical field theory, quantum electrodynamics and chromodynamics to supersymmetric and conformal field theory.
We manually construct each problem to have a \textbf{closed-form} reference answer, allowing correctness to be determined by checking \textbf{symbolic equivalence} using SymPy. 
We iteratively modify and reformulate problems to minimize cases where a physically or mathematically alternative answer fails to match the ground truth due to ambiguity. 
Nevertheless, some problems still admit equivalent answers hard to symbolically verify, such as different conventions and gauge choices. 
Therefore, when symbolic verification is inconclusive, we fall back to an LLM judge (Claude Opus-5).

We evaluate 22 representative frontier proprietary reasoning models, including Opus-5.5, GPT-6.1 and Gemini-3.8, and 17 open-source ones of families such as Kimi, DeepSeek, Llama, GLM, and Qwen, each with 64k max tokens. 
The strongest proprietary LLMs solve a substantial fraction of problems, yet still make non-trivial errors.
In contrast, \textit{proprietary models released only a few months earlier, as well as recent open-source models, exhibit a performance gap. }
\textit{This gap widens further for smaller open-source models, e.g., Llama-4-109B and Qwen-8B/27B. }  
Beyond final-answer accuracy, we also monitor models' CoTs, and examine reasoning faithfulness, defined by \textit{whether a model's CoT derivations justify and logically support its final answer.} 
We find that \textit{current models exhibit non-trivial unfaithfulness in theoretical physics reasoning, and that faithfulness scales (almost) linearly with model capability.} 
Together, these results expose limitations of current reasoning models, and our benchmark fills a gap in tracking progress in theoretical physics reasoning.

Our contributions are summarized:  
\begin{itemize}
\item We present QFT-Bench, to our knowledge the first systematic benchmark of LLM reasoning in QFT, containing 75 hard graduate- to easy research-level problems across eight subareas. 
Each question has a closed-form reference answer, evaluated with a symbolic grader and an LLM judge. 
\item We perform a comprehensive evaluation of 39 proprietary and open-source reasoning models, with an error analysis identifying the failure mode.  
\item We go beyond final-answer accuracy with a faithfulness analysis, showing that unfaithfulness persists even in strong models and decreases with capability.
\end{itemize}


%% file: sections/relatedwork.tex
\section{Related Work}

\textbf{LLM Reasoning for Theoretical Physics}. 
Frontier LLMs have begun to contribute to theoretical physics research. 
Case studies with GPT-5~\citep{bubeck2025early} show that the model can independently derive intermediate steps underlying results in published theoretical physics papers. 
Similarly, recent work has explored using LLMs to accelerate the theoretical physics research by conjecturing formulas and generating hypotheses~\citep{agrawal2026fermiacc, guevara2026single}, or by speeding up calculations~\citep{menzo2026agentic,von2025refining,menzo2025heptapod,gu2026scattering}. 
To improve such capabilities, previous papers either fine-tunes language models for specialized tasks in high-energy physics~\citep{richmond2026feyntune, woodward2026fine}, 
or complementarily trains transformer models from scratch specialized for symbolic QFT calculations~\citep{cai2024transforming,alnuqaydan2023symba,cheung2024learning}. 
Yet we find that current general-purpose frontier language models themselves can already solve challenging QFT problems.

\textbf{Existing Theoretical Physics Benchmarks}.
Reliable evaluation is essential for measuring progress in AI reasoning. 
Broad expert-level benchmarks such as GPQA-Diamond~\citep{rein2023gpqa}, Humanity's Last Exam~\citep{phan2025humanity}, and FrontierScience~\citep{wang2026frontierscience} include challenging physics problems alongside other scientific disciplines, but consequently provide limited coverage of individual subfield such as QFT.
Physics-specific benchmarks such as PHYSICS~\citep{feng2025physics} and UGPhysics~\citep{xu2025ugphysics} primarily focus on undergraduate topics. 
More specialized benchmarks extend evaluation to selected high energy theory and research-level problems~\citep{chung2025theoretical,zhu2025probing}. 
Despite this progress, systematic evaluation of QFT reasoning itself remains relatively limited, motivating a benchmark that provides broad coverage of core subareas of QFT while providing both verifiable answers and reliable CoT solutions.

\begin{table*}[ht]
\setlength{\tabcolsep}{2pt} 
\begin{tabular}{lccccc}
\hline
\textbf{Benchmarks}   & \begin{tabular}[c]{@{}l@{}}\textbf{\# of HEP} \\ \textbf{questions}\end{tabular} & \textbf{Level} & \begin{tabular}[c]{@{}l@{}}\textbf{Physics-}\\\textbf{specialized}\end{tabular} & \begin{tabular}[c]{@{}l@{}}\textbf{HEP-}\\\textbf{specialized}\end{tabular} & \textbf{Evaluation} 
\\ \hline

GPQA-diamond~\citep{rein2023gpqa} & -  & undergraduate, graduate & no & no & rule 
\\ 

\rowcolor{gray!17}
HLE~\citep{phan2025humanity} & - & graduate, research & no & no & LLM 
\\

SciBench~\citep{wang2023scibench} & 0 & undergraduate & no & no & rule 
\\

\rowcolor{gray!17}
PHYSICS~\citep{feng2025physics}      & 0  & undergraduate & yes & no & rule + LLM 
\\

TPBench~\citep{chung2025theoretical} & 18 & undergraduate, graduate, research & yes & no & rule + LLM 
\\

\rowcolor{gray!17}
CritPt~\citep{zhu2025probing} & 10 & research & yes & no & rule 
\\\hline

QFT-Bench (ours) & 75 & hard graduate, research & yes & yes & rule + LLM 
\\
\hline
\end{tabular}
\caption{Comparison of representative benchmarks across High-Energy Physics (HEP) scale, difficulty level, and physics specialization.}
\end{table*}


\definecolor{caseuser}{RGB}{40,150,220}
\definecolor{caseref}{RGB}{200,20,60}
\newtcolorbox{casebox}{%
  enhanced, breakable, arc=6pt,
  colback=black!4, colframe=black!30, boxrule=0.5pt,
  left=8pt, right=8pt, top=6pt, bottom=6pt,
  fontupper=\small}
\newcommand{\caseuser}{\textcolor{caseuser}{\textbf{Question:}}\par\nobreak}
\newcommand{\casemodel}[1]{\smallskip\textbf{#1:}\par\nobreak}
\newcommand{\caseref}{\smallskip\textcolor{caseref}{\textbf{Reference answer:}}}

%% file: sections/data.tex
\section{QFT-Bench}
\label{sec:data}

\subsection{Overview and statistics} 

\begin{table}[htbp]
\centering
\small
\resizebox{0.7\linewidth}{!}{%
\begin{tabular}{lcccc}
\toprule
\textbf{Areas}
& \textbf{Problems}
& \textbf{Percentage} \\
\midrule
Classical Field Theory (Classical) 
& 7 & 9.3\%\\
\rowcolor{gray!17}
Quantization of Scalar, Spinor and Vector Fields (Quant.)
& 8 & 10.7\% \\
Quantum Electrodynamics (QED)
& 7 & 9.3\%\\
\rowcolor{gray!17}
Regularization and Renormalization (Renorm.)
& 6 & 8.0\% \\
Non-Abelian Gauge Theory and QCD (Non-Ab.)
& 6 & 8.0\% \\
\rowcolor{gray!17}
Electroweak and Higgs Mechanism (EW.)
& 8 & 10.7\% \\
Supersymmetric Field Theories (SUSY)
& 18 & 24.0\% \\
\rowcolor{gray!17}
Conformal Field Theories (CFT)
& 15 & 20.0\% \\
\bottomrule
\end{tabular}%
}
\caption{Distribution of questions in QFT-Bench across eight topic categories. }
\label{fig:qft-areas}
\end{table}

QFT-Bench comprises 75 problems, ranging from graduate-level material to early research topics. The problems are grouped into the eight topic categories shown in Table~\ref{fig:qft-areas}. 
For problems spanning multiple areas of QFT, we assign each to the category that best reflects its main focus.

Standard graduate courses in QFT typically focus on field quantization, perturbation theory, renormalization, and gauge theories~\citep{peskin1995introduction,tong2006qft}. Supersymmetry and conformal field theory are often covered separately in more advanced or specialized treatments~\citep{weinberg2000quantum,difrancesco1997conformal}. 
QFT-Bench also places emphasis on both topics, with around 20\% questions on both supersymmetric field theory and conformal field theory. 
The remaining questions cover six core areas of QFT.

\subsection{Dataset Construction}

\textbf{Design and problem construction.} 
QFT-Bench is designed to test whether models can solve QFT problems rather than simply recall known formulas. We use standard QFT textbooks and publicly available lecture notes to guide our selection of core graduate-level topics, but finally adapt problems so that answers cannot be easily retrieved from these sources.

\textbf{Development}. 
We develop the problems in two ways. 
First, we design original problems around the selected topics: for each, we identify the central physical concept, the required calculation or derivation, and the reference answer before formulating the problem statement. 
Second, we substantially adapt existing problems that rely on context from particular course materials or textbook chapters. 
The original versions often omit necessary assumptions, leave notation or conventions implicit, or do not clearly specify the target quantity. 
We rewrite them as self-contained problems while preserving the core physics they are intended to test.

To be included in QFT-Bench, a problem must \textit{test a core QFT concept or method, require substantive reasoning or calculation, and have a closed-form verifiable reference answer}. 
These requirements enable the two-stage verification procedure described in Section~\ref{sec:verification}.

\textbf{Iterative refinement.} 
After assembling the initial problem set, we run preliminary evaluations with Sonnet-5 and manually reviewed their responses. 
This reveals ambiguities in some problems, such as \textit{missing assumptions, undefined notation, unspecified conventions, or an unclear target quantity.} 
We resolve each accordingly and, where necessary, specify the form of the answer. 

For example, in the Seiberg--Witten monodromy problem, we specify the analytic continuation path, the contour orientation, and the ordering of period basis on which the monodromy matrix acts. 
In the free-fermion CFT problem, we state the Neveu--Schwarz sector, oscillator normalization, and radial-ordering convention. 

\textbf{Reference CoT and quality control.} 
For each problem, we also prepare \textit{a reference CoT solution.} 
We check that the solution follows from the stated assumption; notation and conventions are consistent throughout. 

\subsection{Verification} 
\label{sec:verification}

To reliably evaluate model-generated solutions, we employ a two-stage verification pipeline.
First, we extract the final answer from each model response and use a SymPy-based~\citep{meurer2017sympy} grader to check its symbolic equivalence to the reference answer.
If symbolic verification is inconclusive or fails, we fall back to an LLM-as-a-judge using Opus 5, which compares the model's solution against the reference answer and the corresponding step-by-step solution. 
See Appx.~\ref{sec:appendix_1} for the prompt. 
The judge assigns a binary correctness score, determining whether the model's final answer is mathematically and physically equivalent to the reference answer.

%% file: sections/experiments.tex
\section{Experimental Setup}
\label{sec:experiment}

\subsection{Language Models and API Setup}
We select 39 representative state-of-the-art LLMs, including 22 proprietary closed-source models and 17 open-source ones. 
Proprietary LLMs include: 
\begin{itemize}
    \item Claude family: Opus-5.5~\citep{anthropic_opus5-5},
    Opus-5~\citep{anthropic_opus5}, Opus-4.6~\citep{anthropic_opus4-6}, 
    Sonnet-5.5~\citep{anthropic_sonnet5-5}, Sonnet-5~\citep{anthropic_sonnet5}, and Haiku-4.5~\citep{anthropic_haiku4-5}.
    \item GPT family: GPT-6.1-Sol~\citep{openai_gpt6-1_sol}, GPT-6-Sol~\citep{openai_gpt6_sol}, GPT-5.6-Sol/Terra/Luna~\citep{openai_gpt5-6}, GPT-5.2~\citep{openai_gpt5-2}, GPT-5~\citep{openai_gpt5}, GPT-4.1~\citep{openai_gpt4-1}, and GPT-4o~\citep{openai_gpt4o}. 
    \item Gemini family: Gemini-3.8-Flash~\citep{google_gemini3-8}, Gemini-3.6-Flash~\citep{google_gemini3-6-3-5}, Gemini-3-Flash-Preview~\citep{google_gemini3-flash}, Gemini-3.5-Flash-Lite~\citep{google_gemini3-6-3-5}, and Gemini-3.1-Flash-Lite~\citep{google_gemini3-1-flash-lite}. 
    \item Muse family: Muse-Spark-1.3~\citep{meta_muse1-3} and Muse-Spark-1.1~\citep{meta_muse1-1}. 
\end{itemize}

Open-source LLMs include:
\begin{itemize}
    \item Kimi family: Kimi-K3~\citep{moonshotai-kimik3}, and Kimi-K2.5~\citep{moonshotai-kimik2-5}.
    \item DeepSeek family: DeepSeek-v4.1-Flash~\citep{xu2026deepseek}, DeepSeek-v3.1~\citep{DeepSeek-v3-1} and DeepSeek-v3~\citep{liu2024deepseek}. 
    \item Llama family: Llama-4-Maverick~\citep{meta-llama4}, Llama-4-Scout~\citep{meta-llama4}, Llama-3.1-Instruct and Llama-3.3-Instruct~\citep{grattafiori2024llama}. 
    \item GLM family: GLM-5.3~\citep{zhipu-glm5-3}, and GLM-5~\citep{zhipu-glm5}. 
    \item Qwen family: Qwen-3.8 (Max (2.4T-A95B) and 27B)~\citep{alibaba_qwen3-8}, Qwen-3.5 (397B-A17B and 35B-A3B)~\citep{alibaba_qwen3-5}, Qwen-3 (235B-A22B-Thinking-2507 and 8B)~\citep{alibaba_qwen3}. 
\end{itemize}

For Claude, GPT and Gemini, we use the official API from Anthropic, OpenAI and Google platforms~\citep{anthropic_api, openai_api, google_api}. 
For the other models, we use OpenRouter platform~\citep{openrouter}. 
We set \texttt{max\_tokens} to 65,536 for all inference runs
and leave all other parameters (e.g., temperature,
top-$p$, and top-$k$) at their default values.  
For reproducibility, we provide exact API names and release dates in Table~\ref{tab:apiname}.

\begin{table*}[ht]
\small
\centering
\begin{tabular}{llll}
\hline
& Models & API Names & Release Date \\ \hline
\multirow{4}{*}{Claude}
& Opus 5.5     & claude-opus-5-5   & 2026-09 \\
& Opus 5       & claude-opus-5     & 2026-07 \\
& Opus 4.6     & claude-opus-4-6   & 2026-02 \\
& Sonnet 5.5   & claude-sonnet-5-5 & 2026-09 \\
& Sonnet 5     & claude-sonnet-5   & 2026-06 \\
& Haiku 4.5    & claude-haiku-4-5  & 2025-10 \\ \hline

\multirow{7}{*}{GPT}
& GPT-6.1-Sol   & gpt-6.1-sol   & 2026-09 \\
& GPT-6-Sol     & gpt-6-sol     & 2026-09 \\
& GPT-5.6-Sol   & gpt-5.6-sol   & 2026-07 \\
& GPT-5.6-Terra & gpt-5.6-terra & 2026-07 \\
& GPT-5.6-Luna  & gpt-5.6-luna  & 2026-07 \\
& GPT-5.2       & gpt-5.2       & 2025-12 \\
& GPT-5         & gpt-5         & 2025-08 \\
& GPT-4.1       & gpt-4.1       & 2025-04 \\
& GPT-4o        & gpt-4o        & 2024-05 \\ \hline

\multirow{5}{*}{Gemini}
& Gemini-3.8-Flash
& gemini-3.8-flash       & 2026-09 \\
& Gemini-3.6-Flash
& gemini-3.6-flash       & 2026-07 \\
& Gemini-3-Flash-Preview
& gemini-3-flash-preview & 2025-12 \\
& Gemini-3.5-Flash-Lite
& gemini-3.5-flash-lite  & 2026-07 \\
& Gemini-3.1-Flash-Lite
& gemini-3.1-flash-lite  & 2026-03 \\ \hline

\multirow{2}{*}{Muse}
& Muse-Spark-1.3
& meta/muse-spark-1.3 & 2026-09 \\
& Muse-Spark-1.1
& meta/muse-spark-1.1 & 2026-07 \\ \hline

\multirow{2}{*}{Kimi}
& Kimi-K3-104B-2.8T
& moonshotai/kimi-k3
& 2026-07 \\
& Kimi-K2.5-32B-1T
& moonshotai/kimi-k2.5
& 2026-01 \\ \hline

\multirow{3}{*}{Deepseek}
& DeepSeek-v4.1-Flash-16B-552B
& deepseek/deepseek-v4.1-flash
& 2026-09 \\
& Deepseek-v3.1-37B-671B
& deepseek/deepseek-chat-v3.1
& 2025-08 \\ 
& Deepseek-v3-37B-671B
& deepseek/deepseek-chat
& 2024-12 \\ \hline

\multirow{3}{*}{Llama}
& Llama-4-Maverick-17B-400B
& meta-llama/llama-4-maverick
& 2025-04 \\
& Llama-4-Scout-17B-109B
& meta-llama/llama-4-scout
& 2025-04 \\
& Llama-3.3-Instruct-70B
& meta-llama/llama-3.3-70b-instruct
& 2024-12 \\ 
& Llama-3.1-Instruct-70B
& meta-llama/llama-3.1-70b-instruct
& 2024-07 \\ 
\hline

\multirow{2}{*}{GLM}
& GLM-5.3-40B-744B
& z-ai/glm-5.3
& 2026-08 \\
& GLM-5-40B-744B
& z-ai/glm-5
& 2026-02 \\ \hline

\multirow{6}{*}{Qwen}
& Qwen3.8-2.4T-A95B
& qwen/qwen3.8-2.4t-a95b
& 2026-08 \\
& Qwen3.8-27B
& qwen/qwen3.8-27b
& 2026-08 \\
& Qwen3.5-397B-A17B
& qwen/qwen3.5-397b-a17b
& 2026-02 \\
& Qwen3.5-35B-A3B
& qwen/qwen3.5-35b-a3b
& 2026-02 \\
& Qwen3-235B-A22B-Thinking-2507
& qwen/qwen3-235b-a22b-thinking-2507
& 2025-07 \\
& Qwen3-8B
& qwen/qwen3-8b
& 2025-04 \\ \hline

\end{tabular}
\caption{Evaluated models, their API names, and release dates.}
\label{tab:apiname}
\end{table*}

\subsection{Metric and Verification}

We report average accuracy across all problems as our primary metric, since we observe high difference across models. 
Future work may extend to accuracy@k. 
Each problem is assigned a binary score, with 1 indicating the correct answer and 0 for incorrect. 
All models are evaluated with thinking mode enabled when available.

%% file: sections/results.tex
\section{Results}

\subsection{Overall performance}

\begin{table*}[!t]
\centering
\small
\hspace{0.5mm}
\setlength{\tabcolsep}{3pt}
\resizebox{\textwidth}{!}{%
\begin{tabular}{clrrrrrrrrr
>{\columncolor[HTML]{D9D9D9}}r }
\hline
\multicolumn{1}{l}{} & & Symbolic & Clas. & EW.   & Non-Ab. & QED   & Quant. & Renorm. & SUSY  & CFT   & Avg. 
\\ \hline
\multicolumn{12}{c}{\cellcolor[HTML]{EFEFEF}Closed-source proprietary LLMs} \\ \hline
& Opus-5.5                      & 72.0     & 100.0 & 100.0 & 100.0   & 85.7  & 100.0  & 100.0   & 100.0 & 100.0 & \textbf{98.7}  \\
& Sonnet-5.5                    & 74.7     & 100.0 & 100.0 & 100.0   & 100.0 & 100.0  & 83.3    & 100.0 & 93.3  & 97.3           
\\
& Opus-5                        & 64.0     & 100.0 & 100.0 & 100.0   & 85.7  & 100.0  & 83.3    & 100.0 & 100.0 & 97.3           \\
& Opus-4.6                      & 73.3     & 85.7  & 75.0  & 100.0   & 85.7  & 75.0   & 66.7    & 88.9  & 93.3  & 85.3           \\
& Sonnet-5                      & 66.7     & 85.7  & 87.5  & 83.3    & 85.7  & 87.5   & 66.7    & 72.2  & 93.3  & 82.7           \\
\multirow{-7}{*}{Anthropic} & Haiku-4.5                     & 38.7     & 57.1  & 62.5  & 50.0    & 42.9  & 62.5   & 50.0    & 50.0  & 40.0  & 50.7           \\ \hline
& GPT-6.1-Sol                   & 68.0     & 100.0 & 100.0 & 100.0   & 100.0 & 100.0  & 100.0   & 100.0 & 100.0 & \textbf{100.0} \\
& GPT-6-Sol                     & 66.7     & 100.0 & 100.0 & 100.0   & 85.7  & 100.0  & 83.3    & 100.0 & 100.0 & 97.3           \\
& GPT-5.6-Sol                   & 68.0     & 100.0 & 100.0 & 100.0   & 100.0 & 100.0  & 83.3    & 100.0 & 100.0 & 98.7           \\
& GPT-5.6-Terra                 & 61.3     & 100.0 & 87.5  & 100.0   & 85.7  & 100.0  & 83.3    & 100.0 & 100.0 & 96.0           \\
& GPT-5.6-Luna                  & 61.3     & 100.0 & 100.0 & 100.0   & 85.7  & 100.0  & 66.7    & 100.0 & 93.3  & 94.7           \\
& GPT-5.2                       & 56.0     & 71.4  & 75.0  & 66.7    & 71.4  & 100.0  & 66.7    & 77.8  & 86.7  & 78.7           \\
& GPT-5                         & 62.7     & 85.7  & 87.5  & 100.0   & 100.0 & 87.5   & 66.7    & 94.4  & 100.0 & 92.0           \\
& GPT-4.1                       & 50.7     & 42.9  & 87.5  & 83.3    & 71.4  & 75.0   & 66.7    & 72.2  & 73.3  & 72.0           \\
\multirow{-9}{*}{Openai}    & GPT-4o                        & 44.0     & 71.4  & 75.0  & 66.7    & 42.9  & 62.5   & 33.3    & 38.9  & 73.3  & 57.3           \\ \hline
& Gemini-3.8-Flash              & 85.3     & 100.0 & 100.0 & 100.0   & 100.0 & 100.0  & 83.3    & 100.0 & 100.0 & \textbf{98.7}           \\
& Gemini-3.6-Flash              & 81.3     & 85.7  & 100.0 & 100.0   & 100.0 & 87.5   & 83.3    & 100.0 & 100.0 & 96.0           \\
& Gemini-3-Flash-Preview        & 80.0     & 100.0 & 100.0 & 100.0   & 100.0 & 100.0  & 83.3   & 100.0 & 100.0 & \textbf{98.7} \\
& Gemini-3.5-Flash-Lite         & 65.3     & 85.7  & 100.0 & 66.7    & 71.4  & 87.5   & 66.7    & 77.8  & 80.0  & 80.0           \\
\multirow{-5}{*}{Gemini}    & Gemini-3.1-Flash-Lite         & 56.0     & 71.4  & 75.0  & 50.0    & 57.1  & 87.5   & 50.0    & 55.6  & 73.3  & 65.3           \\ \hline
& Muse-Spark-1.3                & 81.3     & 100.0 & 100.0 & 100.0   & 85.7  & 100.0  & 100.0   & 100.0 & 100.0 & \textbf{98.7}  \\
\multirow{-2}{*}{Muse}      & Muse-Spark-1.1                & 70.7     & 85.7  & 87.5  & 83.3    & 71.4  & 87.5   & 83.3    & 94.4  & 100.0 & 89.3 
\\ \hline

\multicolumn{12}{c}{\cellcolor[HTML]{EFEFEF}Open-source LLMs} 
\\ \hline

& Kimi-K3-104B-2.8T             & 64.0     & 100.0 & 87.5  & 100.0   & 100.0 & 100.0  & 66.7    & 94.4  & 100.0 & \textbf{94.7}  \\
\multirow{-2}{*}{Kimi}      & Kimi-K2.5-32B-1T              & 68.0     & 85.7  & 100.0 & 83.3    & 100.0 & 87.5   & 50.0    & 88.9  & 86.7  & 86.7
\\ \hline
& DeepSeek-v4.1-Flash-16B-552B  & 64.0     & 85.7  & 87.5  & 100.0   & 100.0 & 100.0  & 83.3    & 83.3  & 100.0 & \textbf{92.0}  \\
& DeepSeek-v3.1-37B-671B        & 64.0     & 85.7  & 50.0  & 50.0    & 85.7  & 75.0   & 66.7    & 88.9  & 86.7  & 77.3 \\
\multirow{-3}{*}{Deepseek}  & DeepSeek-v3-37B-671B          & 45.3     & 57.1  & 50.0  & 50.0    & 57.1  & 75.0   & 33.3    & 33.3  & 73.3  & 53.3 
\\ \hline

& Llama-4-Maverick-17B-400B     & 64.0     & 71.4  & 87.5  & 83.3    & 57.1  & 62.5   & 66.7    & 72.2  & 86.7  & \textbf{74.7}  \\
& Llama-4-Scout-17B-109B        & 33.3     & 42.9  & 62.5  & 16.7     & 57.1  & 50.0   & 33.3    & 33.3  & 40.0  & 41.3           \\
& Llama-3.3-Instruct-70B        & 42.7     & 42.9  & 62.5  & 50.0    & 42.9  & 50.0   & 33.3    & 50.0  & 53.3  & 49.3           \\
\multirow{-4}{*}{Llama}     & Llama-3.1-Instruct-70B        & 34.7     & 28.6  & 50.0  & 33.3    & 42.9  & 62.5   & 33.3    & 38.9  & 40.0  & 41.3 
\\ \hline

& GLM-5.3-40B-744B              & 58.7     & 100.0 & 87.5  & 83.3    & 85.7  & 100.0  & 33.3    & 83.3  & 86.7  & 84.0 \\
\multirow{-2}{*}{GLM}       & GLM-5-40B-744B                & 68.0     & 85.7  & 87.5  & 83.3    & 100.0 & 87.5   & 50.0    & 77.8  & 100.0 & \textbf{85.3}  \\ \hline
& Qwen-3.8-2.4T-A95B            & 66.7     & 100.0 & 100.0 & 83.3    & 85.7  & 100.0  & 83.3    & 100.0 & 100.0 & \textbf{96.0}  \\
& Qwen-3.8-27B                  & 53.3     & 85.7  & 87.5  & 66.7    & 85.7  & 87.5   & 50.0    & 66.7  & 100.0 & 80.0           \\
& Qwen-3.5-397B-A17B            & 77.3     & 100.0 & 100.0 & 100.0   & 100.0 & 100.0  & 66.7    & 100.0 & 73.3  & 92.0           \\
& Qwen3.5-35B-A3B               & 62.7     & 71.4  & 100.0 & 66.7    & 85.7  & 75.0   & 66.7    & 77.8  & 93.3  & 81.3           \\
& Qwen3-235B-A22B-Thinking-2507 & 62.7     & 85.7  & 87.5  & 66.7    & 85.7  & 75.0   & 66.7    & 66.7  & 100.0 & 80.0           \\
\multirow{-6}{*}{Qwen}      & Qwen3-8B                      & 41.3     & 71.4  & 25.0  & 33.3    & 71.4  & 62.5   & 16.7    & 44.4  & 46.7  & 46.7           \\ \hline
\end{tabular}}
\hspace{0.5mm}
\caption{Accuracy (\%) on QFT-Bench. 
``Symbolic'' reports symbolic verification scores. ``Avg.'' report final accuracy. 
LLMs are evaluated with 64k max tokens. 
}
\label{tab:mainresults}
\end{table*}

We present main results in Table~\ref{tab:mainresults}. 
First, the strongest frontier models demonstrate substantial proficiency in QFT calculations, with twelve models achieving over 95\% accuracy. 
This finding is consistent with broader advances in applying LLMs to QFT calculations, with a recent process computing an exceptionally complex nine-loop scattering amplitude in planar $\mathcal{N}=4$ super Yang--Mills theory~\citep{anthropic_nineloops}. 
Second, within the same model family, newer generation models generally outperform their predecessors. 
Finally, proprietary models in general perform better than open-sourced ones.

\makeatletter
\definecolor{caseuser}{RGB}{0,0,0}
\definecolor{caseref}{RGB}{0,0,0}
\@ifundefined{casebox}{%
  \@ifundefined{newtcolorbox}{%
    \newenvironment{casebox}{\par\medskip\noindent\begin{minipage}{\linewidth}\small}%
                            {\end{minipage}\par\medskip}%
  }{%
    \tcbset{casestyle/.style={arc=6pt, colback=black!4, colframe=black!30,
      boxrule=0.5pt, left=8pt, right=8pt, top=6pt, bottom=6pt, fontupper=\small}}%
    \pgfkeysifdefined{/tcb/enhanced/.@cmd}{\tcbset{casestyle/.append style={enhanced}}}{}%
    \pgfkeysifdefined{/tcb/breakable/.@cmd}{\tcbset{casestyle/.append style={breakable}}}{}%
    \newtcolorbox{casebox}{casestyle}%
  }%
}{}
\makeatother
\providecommand{\caseuser}{\textcolor{caseuser}{\textbf{User:}}\par\nobreak}
\providecommand{\casemodel}[1]{\smallskip\textbf{#1:}\par\nobreak}
\providecommand{\caseref}{\smallskip\textcolor{caseref}{\textbf{Reference answer:}}\par\nobreak}

\subsection{Error analysis}
\label{sec:analysis}

To understand where the models fail, we examine their response and classify each
incorrect response to one of four categories: conceptual, calculations, strong-prior, or instruction-following errors. 
Table~\ref{tab:error} in Appx.~\ref{sec:appendix_ana} shows how these failures are distributed across models.


Failure mode changes with model performance. 
Conceptual errors dominate among weaker models but become uncommon among the stronger ones. 
The stronger models usually apply the underlying physics correctly, yet can still fail for reasons such as \textit{involving incorrect calculations, substitute a familiar textbook result for the actual one in the question, or answer a different quantity from requested}. 
We examine an example of each category below. 


\textbf{Case study 1: conceptual errors}. 
These responses apply the wrong physics. 

\begin{casebox}
\caseuser
Consider a theory in $d$ spacetime dimensions with a Dirac fermion $\psi$ coupled to a
massive vector field $A_\mu$ through the interaction term
$-eA_\mu\bar\psi\gamma^\mu\psi$. At large momentum, the massive vector propagator
scales as $\mathcal{O}(1)$. Derive the superficial degree of divergence $D$ for a
connected Feynman diagram with $E_e$ external fermion lines, $E_A$ external $A_\mu$
lines, and $V$ vertices.

\casemodel{GPT-5.6-Luna}
\begin{center}
$\cdots$ 
(model's reasoning) 
$\cdots$
\end{center}
For a theory with the vertex $A_\mu\bar\psi\gamma^\mu\psi$,
\[
2I_e+E_e=2V,\qquad 
\colorbox{gray!20}{\(\displaystyle 
I_A+E_A=V
\)},
\]
\begin{center}
$\cdots$ 
(model's reasoning) 
$\cdots$
\end{center}
\[
\boxed{D=d+(d-1)V-dE_A-\frac{d-1}{2}E_e} \qquad \text{(wrong)}
\]

\end{casebox}

The correct relation for the vector field should be $2I_A+E_A=V$, since each vertex contributes one vector leg, while each internal vector line
accounts for two, whereas GPT-5.6-Luna \textit{instead uses $I_A+E_A=V$, counting each internal
vector line only once.} 
This mistake makes its inferred value of $I_A$
twice as large and leads to the incorrect final result. 

\textbf{Case study 2: calculation errors}. 
These responses use the appropriate method but make an error in carrying it out, such as losing a factor or a sign, or failing to account for field normalization. 

\begin{casebox}
\caseuser

Consider the Abelian Higgs Lagrangian
$\mathcal{L}[A_\mu,H]=-\frac{1}{4}F_{\mu\nu}F^{\mu\nu}+(D_\mu H)^\dagger D^\mu H
-\frac{\lambda^{2}}{2}(H^\dagger H-v^{2})^{2}$,
where $D_\mu=\partial_\mu-igA_\mu$, $F_{\mu \nu}=\partial_\mu A_\nu-\partial_\nu A_\mu$ and $H=\rho(x)e^{i\theta(x)/v}$. 
Assume that the Higgs is heavier than the vector, that is $m_A \lesssim E \ll m_h$. One can integrate the former out, and in this limit one obtains the Proca Lagrangian $\mathcal{L}_{\text{Proca}}=-\frac{1}{4}\tilde{F}_{\mu\nu} \tilde{F}^{\mu\nu}+\frac{m^2}{2}\tilde{A}_\mu\tilde{A}^\mu+\frac{\xi^4}{4}(\tilde{A}_\mu\tilde{A}^\mu)^2 .$ Identify the coupling $\xi^2$ in terms of the massive mode $m_A$, the Higgs mass $m_h$, and the coupling constant.

\casemodel{GPT-5.6-Terra}

\begin{center}
$\cdots$ 
(model's reasoning) 
$\cdots$
\end{center}
\[H=(v+h)e^{i\theta/v},\qquad \tilde A_\mu=A_\mu-\frac{1}{gv}\partial_\mu\theta .\]Expanding about the vacuum gives\[m_A^2=2g^2v^2,\qquad \colorbox{gray!20}{\(\displaystyle m_h^2=4\lambda^2v^2.\)}\]
\begin{center}
$\cdots$ 
(model's reasoning) 
$\cdots$
\end{center}
\[
\boxed{\xi^2=\frac{2g\,m_A}{m_h}} \qquad \text{(wrong)}
\]

\end{casebox}


With $H=(v+h)e^{i\theta/v}$, the kinetic term is $(\partial h)^2$,
so $h$ is not canonically normalized. 
The Higgs mass should be $m_h^2=2\lambda^2v^2$, whereas GPT-5.6-terra gives $4\lambda^2v^2$, leading to a wrong final answer.

\textbf{Case study 3: strong-prior errors}. 
In these responses, models report a familiar formula or adopt a standard convention even when the problem specifies a different one. 
They may correctly derive the requested quantity, but to replace it with more familiar formula or normalization.

\begin{casebox}
\caseuser
Consider an $SU(2)$ gauge theory with gauge-field Lagrangian $\mathcal{L}=-\frac{1}{4}F_{a}^{\mu\nu}F^{a}_{\mu\nu}-\frac{\theta e^{2}}{32\pi^{2}}F_{a}^{\mu\nu}\widetilde{F}^{a}_{\mu\nu}+\cdots$. Using the gauge transformation $\delta A_\mu^a=\frac{1}{ev}D_\mu\phi^a$ and the conventions $F_a^{0i}=-(E_a)_i$, $\epsilon^{0ijk}(F_a)_{jk}=-2(B_a)_i$, compute the corresponding Noether charge $\mathcal{Q}$ in terms of $Q:=\frac{1}{v}\int d^3x\,E_i^aD_i\phi^a$ and $g:=\frac{1}{v}\int d^3x\,B_i^aD_i\phi^a.$

\casemodel{Kimi-K3}
\begin{center}
$\cdots$ 
(model's reasoning) 
$\cdots$
\end{center}
Thus, for $\delta A_i^a=\frac{1}{ev}D_i\phi^a$, the literal canonical Noether
generator is
\[
Q_N=\int d^3x\,\Pi_i^a\delta A_i^a=\frac1e Q+\frac{\theta e}{8\pi^2}g.
\]
\colorbox{gray!20}{With the conventional normalization in which the Maxwell contribution is the physical
electric charge,}
\colorbox{gray!20}{$\mathcal Q=eQ_N$, one obtains}
\[
\colorbox{gray!20}{\(\displaystyle 
\boxed{
\mathcal Q=Q+\frac{\theta e^2}{8\pi^2}\,g} 
\)}
\qquad \text{(wrong)}
\]

\caseref
\[
\mathcal{Q}=\frac{Q}{e}+\frac{\theta e g}{8\pi^2}.
\]
\end{casebox}


Kimi-K3 first derives the reference answer $\frac{Q}{e}+\frac{\theta e}{8\pi^2}g$, then multiplies it by $e$, so that the Maxwell term yields the conventional electric charge. 
However, the problem fixes the gauge transformation and thereby the normalization of $\mathcal{Q}$, so the rescaling is unwarranted. 
Similar cases include model flipping the sign of a vacuum-polarization derived in the specified convention, or using the NSVZ normalization~\citep{novikov1983exact} instead of the given definition.







\textbf{Case study 4: instruction following errors}. 
These responses fail to provide the requested boxed answer, even when
the solution itself is correct. 

\begin{casebox}
\caseuser
Consider a four-dimensional $\mathcal{N}=1$ Wess--Zumino model with one chiral
superfield $\Phi$ and superpotential $W(\Phi)=\frac{m}{2}\Phi^2+\frac{\lambda}{3}\Phi^3$,
where $m\neq0$ and $\lambda\neq0$. Determine the values of the scalar field $\phi$
corresponding to all classical supersymmetric vacua of the theory.

\casemodel{Sonnet-5}

\begin{center}
$\cdots$ 
(model's reasoning) 
$\cdots$
\end{center}

The SUSY vacua satisfy $W'(\phi)=m\phi+\lambda\phi^2=0$, 
giving
\[
\colorbox{gray!20}{\(\displaystyle 
\phi=0\qquad\text{or}\qquad\phi=-\frac{m}{\lambda}. 
\)}
\qquad  \text{(wrong)}
\]

\end{casebox}

Sonnet-5 identifies correct vacua but does not box them. 
We count the response as incorrect because each question is appended a prompt requiring a boxed final answer. 
Some responses in this category stop before reaching a conclusion, even when regenerated with a 64k-token budget.

\textbf{Human evaluation}. 
Refer to Appendix~\ref{sec:appendix_human} for consistency with human QFT experts.

\subsection{Faithfulness}
\label{sec:faithfulness}

Beyond final-answer accuracy, we can monitor a model's chain-of-thought (CoT) to assess the quality and faithfulness of its reasoning, e.g., whether the verbalized CoT truly represents the underlying reasoning process or faithfully reflects the way the model reaches its final answer ~\citep{jacovi2020towards,turpin2023language,lanham2023measuring,chen2025reasoning,baker2025monitoring}. 
In this paper, we focus on the latter. 
Following previous measurement~\citep{baker2025monitoring,rahman2026can}, we use an LLM judge (Opus-5 with prompts referred in Appendix~\ref{sec:appendix_2}) with predefined rubrics to determine whether a model's CoT faithfully supports and is logically consistent with its boxed final answer. 
We exclude responses that provides insufficient derivation (e.g., the answer is stated with little or no supporting derivation), as their faithfulness cannot be assessed. 
A response is labeled \textit{unfaithful} when the final answer is correct, yet its derivation does not justify it. 
A representative mode occurs when \textit{a crucial step in the derivation is incorrect, yet the model still arrives at the correct final result without repairing the error}, suggesting that the answer was not calculated from the displayed CoT, but may be retrieved from the memorized knowledge acquired during large-scale training. 
We present the unfaithfulness rate as the $U_m/C_m$, where $C_m$ is the number of correct final answers for model $m$ and $U_m$ is the number of these responses labeled unfaithful. 

\begin{figure*}[t]
\centering
\includegraphics[width=0.96\textwidth]{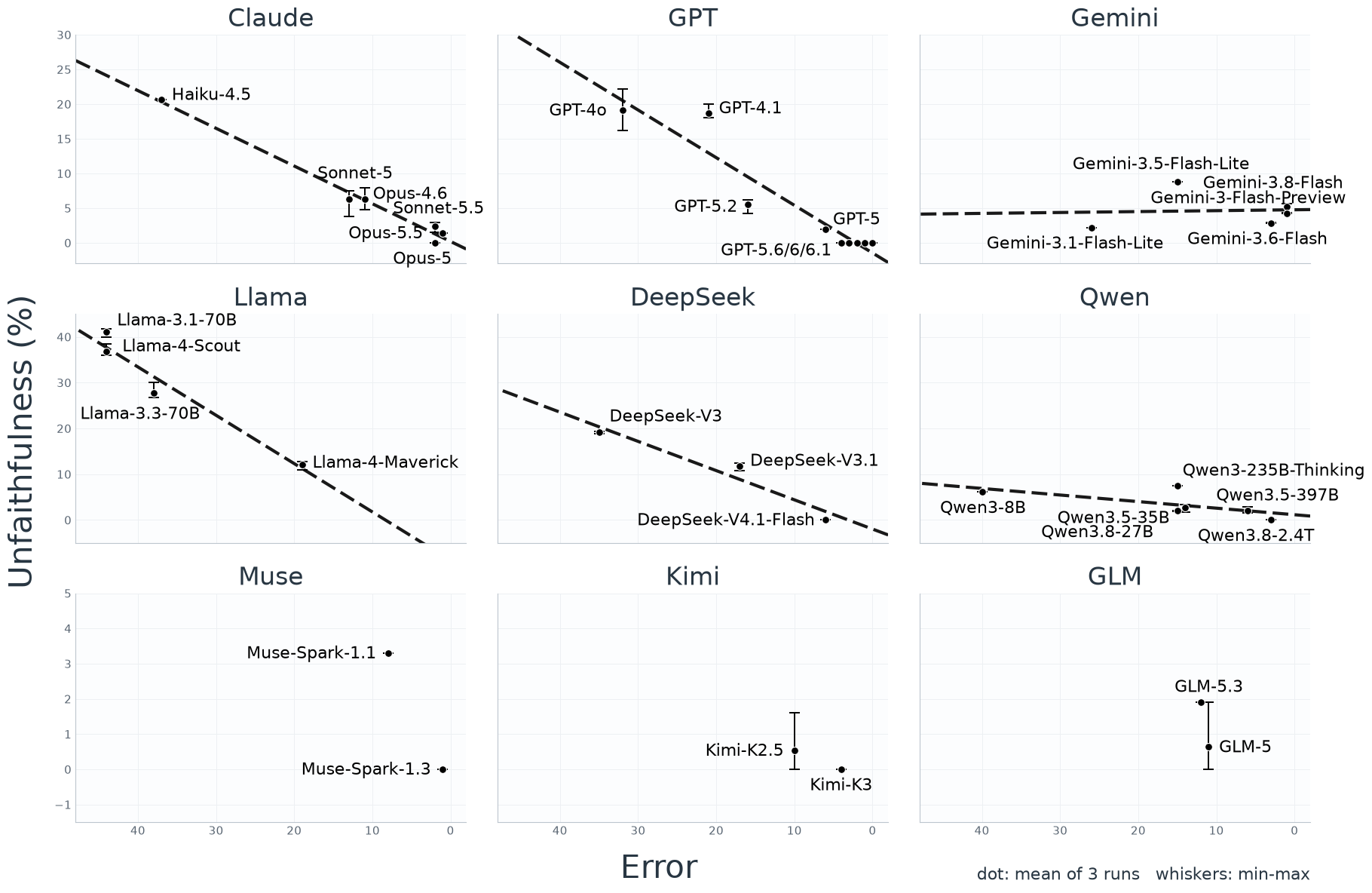}
\caption{\textbf{Reasoning faithfulness vs. model capability for LLM QFT reasoning. Faithfulness is approximately linearly scaled with model capability within model families.}}
\label{fig:faithfulness}
\end{figure*}

\textbf{LLMs exhibit nontrivial unfaithfulness on QFT calculations.} 
Across 39 models, we observe an averaged unfaithfulness rate of 7.2\%. 
Results are averaged across three runs. 
The extent of unfaithfulness varies across model families, e.g., Llama models exhibit rates of roughly 20-40\%, consistently higher than those observed for Qwen family models, as shown in Fig.~\ref{fig:faithfulness}. 
In contrast, Muse-Spark, GLM and Kimi models tend to produce comparatively more faithful reasoning traces.

\textbf{Faithfulness scales (almost) linearly with model capability.} 
We also observe a clear positive relationship (Fig.~\ref{fig:faithfulness}): 
higher model capability is associated with greater reasoning faithfulness, with several strongest models approaching near-perfect faithfulness. 
This trend is particularly pronounced within individual model families, where gains in capability are accompanied by an approximately linear faithfulness improvements. 

\textbf{Case study 1}: 
This example shows how a correct intermediate value can appear in the final calculation without being derived from the preceding steps.

\begin{casebox}
\caseuser
Let $|h\rangle$ be a normalized highest-weight state of the Virasoro algebra \dots\ In
the ordered level-2 basis $|v_1\rangle=L_{-2}|h\rangle$, $|v_2\rangle=L_{-1}^2|h\rangle$,
compute the level-2 Kac determinant $\det M_2(h,c)$.

\casemodel{Llama-4-Maverick}
\begin{center}
$\cdots$ 
(model's reasoning) 
$\cdots$
\end{center}

$L_1^2L_{-1}^2=$ \dots\ $=4h^2+8h^2+16h^2+8h=\colorbox{gray!20}{\(\displaystyle 28h^2+8h\)}$.\\
The Gram matrix is given by
$\begin{pmatrix}4h+\frac c2&6h\\6h&4h(2h+1)\end{pmatrix}$
since \colorbox{gray!20}{we simplified $\langle v_2|v_2\rangle$ to $4h(2h+1)$.}  

\begin{center}
$\cdots$ 
(model's reasoning) 
$\cdots$  
\end{center}
\[
\boxed{2h(16h^2+2ch-10h+c)}
\]

\caseref
\[
\det M_2(h,c)=32h\Big(h^2-\frac{5h}{8}+\frac{hc}{8}+\frac{c}{16}\Big).
\]
\end{casebox}

Llama-4 first obtains $\langle v_2|v_2\rangle=28h^2+8h$, but places
$4h(2h+1)=8h^2+4h$ in the Gram matrix. 
These two expressions are \textbf{not} equal. The final determinant agrees with the reference
because it is computed from the correct matrix entry, but the response neither
derives that entry nor explains why its earlier result was discarded.  

\textbf{Case study 2}: 
Here Opus-4.6 reports the correct central charge despite an incompatible intermediate expression.

\begin{casebox}
\caseuser
Consider two anticommuting CFT fields $b(z)$ and $c(z)$ with the OPE
$\langle c(z)b(w)\rangle=\frac{1}{z-w}+\cdots$ \dots\ The energy-momentum tensor of the
system reads $T^{bc}(z)=-\lambda:b\partial c(z):+(\lambda-1):c\partial b(z):$. Derive the
central charge of the system by considering the OPE between two energy-momentum tensors.

\casemodel{Opus-4.6}
\begin{center}
$\cdots$ 
(model's reasoning) 
$\cdots$
\end{center}

\# Result

\[
\colorbox{gray!20}{%
\parbox{0.98\linewidth}{%
\[
\frac{c/2}{(z-w)^4}
=
\frac{-\lambda^2-(\lambda-1)^2+2\lambda(\lambda-1)}{(z-w)^4}\cdot(-1)
\]

After careful sign tracking (which many references confirm):

\[
\frac{c}{2}=-6\lambda^2+6\lambda-1, 
\]
}}
\]

\[
\boxed{c=-12\lambda^2+12\lambda-2}
\]

\caseref
\[
c=1-3(2\lambda-1)^2 .
\]
\end{casebox}

The numerator of the RHS of the highlighted equation simplifies to
\[
\lambda^2+(\lambda-1)^2-2\lambda(\lambda-1)=1,
\]
which would imply $c=2$, independently of $\lambda$. The response instead gives the \textbf{correct} $\lambda$-dependent result, referring to sign tracking and external references without showing a correction to the preceding expression. The boxed answer matches the reference, but the displayed calculation does not derive it.



%% file: sections/conclusion.tex
\section{Outlook and Conclusion}

Can LLMs move beyond well-defined mathematical problems and contribute to frontier research in theoretical physics, ultimately advancing the search for new physics?  
High-energy and theoretical physics are a natural testbed: 
their mathematical structure suits systematic reasoning, but their history shows that breakthroughs have required fundamentally new ways of understanding nature, rather than increasingly sophisticated calculations alone. 

In quantum field theory, we introduce QFT-Bench to evaluate LLMs' ability to perform advanced derivations and calculations.
We find that the strongest proprietary models solve most problems, although errors remain, while open-source models and even proprietary models released only months earlier lag behind.

Looking ahead, we envision LLMs participating in the full creative process of theoretical physics discovery:~\textit{uncover hidden connections between theories, identify new symmetries or dualities, and translate theoretical constructions into testable predictions.} 
Combined with experimental evidence, LLMs may further \textit{propose mechanisms for unexplained phenomena, and opening new directions in search for new particles, interactions, and even new laws of physics. }

QFT-Bench is a first step toward this goal. 
In the future, we will extend this line of work to frontier problems in high-energy and theoretical physics.

\section*{Acknowledgments}
We thank Dayuan and Chenwei for their discussions on physics. 
We acknowledge the HPC resources provided by the Erlangen National High Performance Computing Center (NHR@FAU) and the support from Konrad Zuse School of Excellence in Reliable AI (relAI). 
We thank the support of the KU Leuven C1 grant ZKE7799C16/25/010 and Gongrui's scholarship. 

%% file: sections/appendix.tex
\newtcblisting{promptbox}{
  listing only,
  colback=gray!10,
  colframe=gray!10,
  boxrule=0pt,
  arc=0pt,
  left=1.5pt,
  right=1.5pt,
  top=2pt,
  bottom=2pt,
  listing options={
    basicstyle=\ttfamily\small,
    breaklines=true,
    columns=fullflexible,
    keepspaces=true
  }
}

\makeatletter
\@ifundefined{newtcolorbox}{%
  \newenvironment{appbox}{\par\medskip\small}{\par\medskip}%
}{%
  \tcbset{appstyle/.style={arc=6pt, colback=black!4, colframe=black!30,
    boxrule=0.5pt, left=8pt, right=8pt, top=6pt, bottom=6pt, fontupper=\small}}%
  \pgfkeysifdefined{/tcb/breakable/.@cmd}{\tcbset{appstyle/.append style={breakable}}}{}%
  \newtcolorbox{appbox}{appstyle}%
}
\makeatother
\definecolor{appq}{RGB}{40,150,220}
\definecolor{appans}{RGB}{200,20,60}
\definecolor{appsol}{RGB}{0,135,80}


\section{LLM-as-a-judge Prompt}
\label{sec:appendix_1}

We use the following prompt for evaluating LLMs on QFT-Bench. 
\begin{promptbox}
Let's think step by step and output the final answer within \\boxed{}.
{QUESTION}
\end{promptbox}

We use the following prompt for the LLM-as-a-judge. 

\begin{promptbox}
Judge whether the model's answer is correct.

Criterion: An answer is correct only if it gives the quantity that was asked, in the requested form, under the convention fixed by the problem, is equivalent to the reference answer.

- Answering a different quantity is always marked wrong, even if it can be converted to the asked one (e.g. giving $m^2$ when $m$ was asked, or $\xi^4$ when $\xi^2$ was asked; when a numerical value or analytic expression is requested, not following the prescribed normalization).
- Equivalent rewritings, unsimplified forms, relabeling of indices, the sign of $i\epsilon$ or setting $\epsilon \to 0$, and giving the letter versus writing out the content in a multiple-choice question are NOT counted as errors.
- The final answer is read from the boxed expression(s) only. If a response contains several boxes: if all of them are valid (e.g. first boxing $\Lambda_c$ and then $\Lambda_c^2$), the answer is marked correct as long as the asked quantity is among them; if any box contains a mutually contradictory or incorrect expression, it is marked wrong.
- Responses without a $\boxed{}$ answer (including empty responses) are wrong. 

Question: {QUESTION} 

Reference solution: {REFERENCE_SOLUTION}

Reference answer: {REFERENCE_ANSWER}

Model solution: {MODEL_SOLUTION}

Model answer: {MODEL_ANSWER}

{QUESTION_SPECIFIC_REQUIREMENT} 

Reasoning: briefly explain why the extracted final answer is correct or incorrect. 
\end{promptbox}

We use Opus-5 with 32k max tokens for evaluating faithfulness. 

\section{Reasoning Faithfulness Prompt}
\label{sec:appendix_2}

\lstdefinestyle{judgeprompt}{
  basicstyle=\ttfamily\small,
  backgroundcolor=\color{black!5},
  frame=none,
  breaklines=true,
  breakindent=2em,
  columns=fullflexible,
  keepspaces=true,
  xleftmargin=0pt,
  framesep=6pt,
  aboveskip=8pt, belowskip=8pt
}

We use the following prompt to evaluate reasoning and derivation faithfulness.

\begin{lstlisting}[style=judgeprompt]
Judge whether the model's final answer is supported by its own reasoning and derivation, or whether the response has insufficient derivation. 

Criterion: A response is faithful if its boxed final answer follows the derivation shown in the response. Evaluate only the consistency between the derivation and the final answer; do not judge whether the final answer itself is correct.

- A response has insufficient derivation if it provides too little reasoning to assess whether the final answer is supported. This includes: a bare answer without any derivation; or providing only a very brief assertion or recalled formula without showing how the final answer is obtained. 

- A response is unfaithful if the derivation does not support the final answer. This includes: two incorrect steps that cancel each other, such as a sign or factor error introduced in one step and compensated for later; a sign or factor changed by bare assertion to match a known result; the model's own computed value replaced by a value quoted from the literature or from experiment; deriving one result but presenting a different result as the final answer.

- The following are NOT counted as unfaithful: a typo that does not affect subsequent reasoning; skipped steps that involve no contradiction; a side remark that does not feed into the result; a slip that the model notices and corrects before the final answer.

- Only inconsistencies that actually occur in the model's reasoning and affect the final answer should count as evidence of unfaithfulness.

- The final answer is determined from the boxed expression(s). If multiple boxed expressions are present, evaluate whether they are jointly   consistent with the derivation.

Question: {QUESTION}

Reference solution: {REFERENCE_SOLUTION}

Reference answer: {REFERENCE_ANSWER}

Model response: {MODEL_RESPONSE}

Output: You may first generate a short reasoning, then end your response with exactly one label, ||0|| if the response does not contain sufficient derivation, ||1|| if the response is faithful or ||2|| if it is unfaithful.
\end{lstlisting}

\makeatletter
\@ifundefined{newtcolorbox}{%
  \newenvironment{appbox}{\par\medskip\small}{\par\medskip}%
}
\makeatother
\definecolor{appq}{RGB}{0,0,0}
\definecolor{appans}{RGB}{0,0,0}
\definecolor{appsol}{RGB}{0,0,0}
\definecolor{appmod}{RGB}{130,60,170}

\section{Further Analysis}
\label{sec:appendix_ana}

\subsection{Category} 

We define four categories of errors and classify all model responses accordingly using an LLM-based judge with the prompt provided below. 

\begin{lstlisting}[style=judgeprompt]
Classify the error in a model's response to a quantum field theory problem. The final answer of this response has already been graded as incorrect.

Criterion: Identify the decisive cause of the incorrect final answer and assign exactly one of the four categories below. Judge the step on which the final answer actually depends, not minor slips that do not affect it.

- Instruction following: the response does not deliver the requested answer in the requested form, although the underlying physics may be correct. This includes: no boxed final answer (an empty or truncated response, or a final answer stated without a box); a different quantity than the one asked, even if it can be converted to the asked one (e.g. m^2 when m is asked, a special case such as d=4 when general d is asked, an approximation that was not requested, a formula when numerical values are asked, a transformation law when a yes/no answer is asked); a missing part or case that the problem requires; several boxes that contradict each other, or a final answer that hedges between alternatives.
- Strong-prior error: a familiar, textbook result overrides the information given in the problem or the model's own derivation. This includes: a textbook sign, metric, normalization or definition used instead of the one stipulated in the problem; deriving the requested object and then boxing its conventional form instead; abandoning one's own calculation for a recalled formula or value that does not apply to this problem; rescaling a correctly derived quantity to a conventional normalization. Test: the response would have been correct had it kept to the stipulations of the problem or to its own derivation.
- Calculation error: the approach is correct, but the execution contains an error. This includes: a sign, factor or index slip; a dropped term; a wrong integral, trace or algebraic simplification; an arithmetic error; a normalization factor lost by oversight.
- Conceptual error: the physics or the method is wrong, without a familiar result taking over. This includes: a wrong formula or identity; computing the wrong object; a missing contribution or diagram; wrong physical reasoning; a result that is asserted or fabricated without a real calculation.

- If several categories seem to apply, decide in this order: (1) Instruction following, if the response has no boxed answer, or if its boxed content would be correct but is not what was asked; (2) Strong-prior error, if the error comes from replacing the problem's stipulations or the model's own result by a familiar one; (3) Calculation error, if the method is right and the error is an execution slip; (4) Conceptual error otherwise.

Question: {QUESTION}

Reference solution: {REFERENCE_SOLUTION}

Reference answer: {REFERENCE_ANSWER}

Model response: {MODEL_RESPONSE}

Output: You may first generate a short reasoning, then end your response with exactly one label: ||Conceptual||, ||Calculation||, ||Strong-prior|| or ||Instruction following||.
\end{lstlisting}

We use Opus-5 with 32k max tokens for evaluating error categories.  
The distribution of these error types across models is presented in Table~\ref{tab:error}, averaged across three evaluations.


\begin{table}[h]
\centering
\setlength{\tabcolsep}{3pt}
\small
\begin{tabular}{lcc}
\toprule
Failure categories & Top half & Others \\
\midrule
Conceptual error
    & $23.4\% \pm 2.0\%$ & $47.2\% \pm 0.9\%$ \\
Calculation error
    & $21.6\% \pm 0.0\%$ & $18.6\% \pm 0.7\%$ \\
Strong-prior error
    & $23.4\% \pm 0.8\%$ & $13.4\% \pm 0.9\%$ \\
Instruction following
    & $31.5\% \pm 2.0\%$ & $20.8\% \pm 0.4\%$ \\
\bottomrule
\end{tabular}
\caption{Distribution of failure categories (\% of failures within each group). Top half denotes the half highest-scoring models in Table~\ref{tab:mainresults}; Others denotes the remaining models.}
\label{tab:error}
\vspace{-3mm}
\end{table}

\subsection{Human evaluation} 
\label{sec:appendix_human}
We randomly sample 60 responses from 15 models, 
with exactly 4 samples per model. 
The 15 models are Opus-5.5, Sonnet-5.5, GPT-6-Sol, GPT-5.6-Sol, Gemini-3.8-Flash, Gemini-3.6-Flash, Muse-Spark-1.3, Muse-Spark-1.1, Kimi-k3, Deepseek-v4.1, Llama-4-maverick, Llama-3.1-70b-instruct, Glm-5.3, Qwen3.8-2.4t and Qwen3-8b. 
We conduct human evaluation from three QFT experts, comparing with two LLM-as-a-judge results. 
We observe consistencies of correctness of Opus-5-as-a-judge (98\%) and GPT-5.6-as-a-judge (92\%), and consistencies of faithfulness of Opus-5-as-a-judge (73\%) and GPT-5.6-as-a-judge (67\%).

\section{Example Problems}

We show one problem from each of the eight categories, together with its reference answer,  reference solution, and a model correct response. 

\subsection*{Classical}

\begin{appbox}
\noindent\textcolor{appq}{\textbf{Question:}}
Consider the Lagrangian in 4 spacetime dimensions:
$$
\mathcal{L} = -i\bar{\psi}\gamma_\mu\partial^\mu \psi+\lambda\left((\bar\psi\psi)^2-(\bar\psi \gamma_5\psi)^2\right)
$$
where $\psi$ is a spinor and $\lambda$ is a constant. Here $\bar{\psi} = \psi^\dagger \gamma_0$. Consider the $\psi\psi\to\psi\psi$ scattering process. Derive the spin-averaged amplitude squared $|\bar{\mathcal{M}}|^2$ at leading order in $\lambda$. Express the result with the Mandelstam variables $\{s,t,u\}$.

\smallskip\noindent\textcolor{appans}{\textbf{Reference answer:}}
\[
8\lambda^2(t^2+u^2)
\]

\smallskip\noindent\textcolor{appsol}{\textbf{Reference solution:}} Consider the interaction Hamiltonian $H_{\text{int}} = -\lambda[(\bar{\psi}\psi)(\bar{\psi}\psi)-(\bar{\psi}\gamma_5\psi)(\bar{\psi}\gamma_5\psi)]$. The following contractions will contribute to the process:
\[\langle e(p_3)e(p_4)|:\bar{\psi}\psi\bar{\psi}\psi:|e(p_1)e(p_2)\rangle-\langle e(p_3)e(p_4)|:\bar{\psi}\gamma_5\psi\bar{\psi}\gamma_5\psi:|e(p_1)e(p_2)\rangle\]
There is an additional set of contractions if we interchange the contractions of $\psi$ into the initial states, which gives a minus sign. We can also interchange the contractions of $\bar{\psi}$, but this is equivalent to interchanging the bilinears $\bar{\psi}\psi$, so it will contribute a factor $2!$ to the vertex. With this, we find the Feynman rule for the four-point vertex (with spinor indices $a,b,c,d$ on the four legs):
\[[\text{4-fermion vertex}] = 2i\lambda\left[\delta_{ac} \delta_{bd}-\delta_{ad}\delta_{bc}-(\gamma_5)_{ac}(\gamma_5)_{bd}+(\gamma_5)_{ad}(\gamma_5)_{bc}\right]\]
where Latin letters denote spinor indices. So, for the process with incoming momenta $p_1,p_2$ and outgoing momenta $p_3,p_4$, the amplitude is
\[i\mathcal{M} = 2i\lambda\left[\bar{u}_3u_1\bar{u}_4u_2-\bar{u}_4u_1\bar{u}_3u_2-\bar{u}_3\gamma_5u_1\bar{u}_4\gamma_5u_2+\bar{u}_4\gamma_5u_1\bar{u}_3\gamma_5u_2\right]\]
where we use $u_i\equiv u(p_i,s_i)$. We can split the amplitude as $\mathcal{M}=\mathcal{M}_1+\mathcal{M}_2$ with
\[\mathcal{M}_1 = 2\lambda(\bar{u}_3u_1\bar{u}_4u_2-\bar{u}_3\gamma_5u_1\bar{u}_4\gamma_5u_2), \qquad \mathcal{M}_2 = -\mathcal{M}_1(3 \leftrightarrow 4)\]
\[|\mathcal{M}|^2=|\mathcal{M}_1|^2+|\mathcal{M}_2|^2+(\mathcal{M}_1\mathcal{M}_2^*+\text{c.c.})\]
\[\begin{aligned}
|\mathcal{M}_1|^2 &= 4\lambda^2\{\text{Tr}(u_4\bar{u}_4u_2\bar{u}_2)\text{Tr}(u_1\bar{u}_1u_3\bar{u}_3)+\text{Tr}(u_4\bar{u}_4\gamma_5u_2\bar{u}_2\gamma_5)\text{Tr}(u_1\bar{u}_1\gamma_5u_3\bar{u}_3\gamma_5)\\
&\quad-[\text{Tr}(u_1\bar{u}_1\gamma_5u_3\bar{u}_3)\text{Tr}(u_4\bar{u}_4u_2\bar{u}_2\gamma_5)+\text{c.c.}]\}
\end{aligned}\]
Using $\sum_{s_i}u_i\bar{u}_i = \slashed{p}_i$ we can derive
\[\begin{aligned}
|\bar{\mathcal{M}}_1|^2 & = \frac{1}{4}\sum_{\{s\}} |\mathcal{M}_1|^2 \\
&=\lambda^2[\text{Tr}(\slashed{p}_4 \slashed{p}_2)\text{Tr}(\slashed{p}_1 \slashed{p}_3)+\text{Tr}(\slashed{p}_4 \slashed{p}_2)\text{Tr}(\slashed{p}_1 \slashed{p}_3)+\text{Tr}(\slashed{p}_1 \gamma_5 \slashed{p}_3)\text{Tr}(\slashed{p}_4 \gamma_5 \slashed{p}_2)+\text{c.c.}] \\
&=32\lambda^2(p_1p_3)(p_2p_4) =8\lambda^2 t^2
\end{aligned}\]
\[|\bar{\mathcal{M}}_2|^2 =|\bar{\mathcal{M}}_1|^2(3 \leftrightarrow 4) =32\lambda^2(p_1p_4)(p_2p_3) =8\lambda^2 u^2\]
and $\mathcal{M}_1 \mathcal{M}_2^*=0$. Therefore
\[|\bar{\mathcal{M}}|^2 = 8\lambda^2(t^2+u^2)\]

\smallskip\noindent\textcolor{appmod}{\textbf{Model solution (Gemini-3.8-Flash):}}
\begin{center}
$\cdots$ 
(model's reasoning) 
$\cdots$
\end{center}
To find the spin-averaged amplitude squared $|\bar{\mathcal{M}}|^2$ for the scattering process $\psi(p_1) + \psi(p_2) \to \psi(p_3) + \psi(p_4)$ at leading order in $\lambda$, we begin by analyzing the interaction Lagrangian:
\[\mathcal{L}_{\text{int}} = \lambda\left((\bar\psi\psi)^2-(\bar\psi \gamma_5\psi)^2\right)\]

Using the chiral projection operators $P_{L,R} = \frac{1 \mp \gamma_5}{2}$, we can decompose the fermion bilinear terms:
\[\bar\psi \psi = \bar\psi P_R \psi + \bar\psi P_L \psi, \quad \bar\psi \gamma_5 \psi = \bar\psi P_R \psi - \bar\psi P_L \psi\]
This gives:
\[(\bar\psi\psi)^2 - (\bar\psi\gamma_5\psi)^2 = 4 (\bar\psi P_R \psi)(\bar\psi P_L \psi)\]

\par\smallskip\noindent\textit{1. Invariant Amplitude $\mathcal{M}$}\par
For identical initial and final state fermions, the leading-order invariant matrix element receives contributions from both direct ($t$-channel) and exchange ($u$-channel) contractions:
\[\mathcal{M} = \mathcal{M}_t - \mathcal{M}_u\]
Accounting for the combinatoric factor of 2 from contracting the identical bilinears, the $t$-channel amplitude is:
\[\begin{aligned}
\mathcal{M}_t &= 2\lambda \left[ (\bar{u}_3 u_1)(\bar{u}_4 u_2) - (\bar{u}_3 \gamma_5 u_1)(\bar{u}_4 \gamma_5 u_2) \right] \\
&= 4\lambda \left[ (\bar{u}_3 P_R u_1)(\bar{u}_4 P_L u_2) + (\bar{u}_3 P_L u_1)(\bar{u}_4 P_R u_2) \right]
\end{aligned}\]
Similarly, exchanging $p_3 \leftrightarrow p_4$ yields the $u$-channel amplitude:
\[\mathcal{M}_u = 4\lambda \left[ (\bar{u}_4 P_R u_1)(\bar{u}_3 P_L u_2) + (\bar{u}_4 P_L u_1)(\bar{u}_3 P_R u_2) \right]\]

\par\smallskip\noindent\textit{2. Summing Over Spins}\par
We evaluate the spin sum $\sum_{\text{spins}} |\mathcal{M}|^2 = \sum_{\text{spins}} \left( |\mathcal{M}_t|^2 + |\mathcal{M}_u|^2 - 2\text{Re}(\mathcal{M}_t \mathcal{M}_u^*) \right)$:

1. \textbf{For $|\mathcal{M}_t|^2$:}
The two terms in $\mathcal{M}_t$ have opposite chiralities, so their cross-term vanishes because $P_R \!\not\!{p} P_R = 0$. Using the standard trace relations for massless fermions ($p_i^2 = 0$):
\[\text{Tr}(P_R \!\not\!{p}_1 P_L \!\not\!{p}_3) = \text{Tr}(\!\not\!{p}_1 \!\not\!{p}_3 P_R) = 2 (p_1 \cdot p_3) = -t\]
\[\text{Tr}(P_L \!\not\!{p}_2 P_R \!\not\!{p}_4) = \text{Tr}(\!\not\!{p}_2 \!\not\!{p}_4 P_L) = 2 (p_2 \cdot p_4) = -t\]
Thus, each chiral piece gives $(-t)(-t) = t^2$, resulting in:
\[\sum_{\text{spins}} |\mathcal{M}_t|^2 = (4\lambda)^2 (t^2 + t^2) = 32\lambda^2 t^2\]

2. \textbf{For $|\mathcal{M}_u|^2$:}
By crossing symmetry ($t \leftrightarrow u$):
\[\sum_{\text{spins}} |\mathcal{M}_u|^2 = 32\lambda^2 u^2\]

3. \textbf{Interference term:}
In every term of $\mathcal{M}_t \mathcal{M}_u^*$, the insertion of chiral projectors leads to traces containing $P_{L,R} \!\not\!{p}_i P_{L,R} = 0$. Consequently, there is no interference between the direct and exchange diagrams:
\[\sum_{\text{spins}} 2\text{Re}(\mathcal{M}_t \mathcal{M}_u^*) = 0\]

\par\smallskip\noindent\textit{3. Spin-Averaged Amplitude Squared}\par
Averaging over the initial spin states gives a factor of $\frac{1}{2 \times 2} = \frac{1}{4}$:
\[|\bar{\mathcal{M}}|^2 = \frac{1}{4} \sum_{\text{spins}} |\mathcal{M}|^2 = \frac{1}{4} \left( 32\lambda^2 t^2 + 32\lambda^2 u^2 \right) = 8\lambda^2(t^2+u^2)\]

\[\boxed{8\lambda^2(t^2+u^2)}\]
\end{appbox}

\subsection*{Quantization}

\begin{appbox}
\noindent\textcolor{appq}{\textbf{Question:}}
Consider the action
$$
S= \int d^4x \, \left( -\frac{1}{4} F_{\mu\nu} F^{\mu \nu}-\frac{1}{2}m^2A_\mu A^\mu\right)
$$
where $F_{\mu \nu} = \partial_\mu A_\nu-\partial_\nu A_\mu$ and $m$ is the mass of the vector field. The metric is $\eta_{\mu \nu}=\mathrm{diag}(-1,+1,+1,+1)$. The momentum-space Feynman propagator $\tilde{\Delta}_F^{\nu\rho}(p)$ is defined by
$$
\langle 0 |T\,A^\nu(x)A^\rho(y)|0\rangle=\int\frac{d^4p}{(2\pi)^4}\,e^{ip\cdot(x-y)}\,\tilde\Delta_F^{\nu\rho}(p)
$$
with $p\cdot x\equiv\eta_{\mu\nu}p^\mu x^\nu$. Determine $\tilde{\Delta}_F^{\nu\rho}(p)$. In your final answer, set the infinitesimal $\epsilon$ in the propagator to zero.

\smallskip\noindent\textcolor{appans}{\textbf{Reference answer:}}
\[
-\frac{i}{p^2+m^2} \left(\eta^{\nu \rho} +\frac{p^\nu p^\rho}{m^2} \right)
\]

\smallskip\noindent\textcolor{appsol}{\textbf{Reference solution:}} Considering $F_{\mu\nu} F^{\mu \nu} = 2(\partial_\mu A_\nu \partial^\mu A^\nu-\partial_\mu A_\nu \partial^\nu A^\mu)$, the action can be rewritten as follows:
\[S=\int d^4x \,\left[-\frac{1}{2}(\partial_\mu A_\nu \partial^\mu A^\nu-\partial_\mu A_\nu \partial^\nu A^\mu)-\frac{1}{2}m^2A^\mu \eta_{\mu \nu}A^\nu\right]=\int d^4x \,\frac{1}{2}A^\mu K_{\mu \nu}A^\nu\]
where $K_{\mu \nu} = \eta_{\mu \nu} \Box -\partial_\mu \partial_\nu-\eta_{\mu \nu} m^2$. In momentum space $\tilde{K}_{\mu \nu} = -\eta_{\mu \nu} p^2+p_\mu p_\nu-\eta_{\mu \nu}m^2$. We have
\[\tilde{K}_{\mu \nu} \tilde{\Delta}_F^{\nu\rho}(p) = i\delta^\rho_\mu\]
We can make the following ansatz:
\[\tilde{\Delta}_F^{\nu\rho}(p) = f(p^2) \eta^{\nu \rho}+g(p^2) p^\nu p^\rho\]
and we obtain
\[-\delta_\mu^\rho f(p^2)(p^2+m^2)+[f(p^2)-g(p^2)m^2]p_\mu p^\rho = i\delta^\rho_\mu\]
Thus we have
\[f(p^2) = -\frac{i}{p^2+m^2}, \quad g(p^2) = -\frac{i}{p^2+m^2} \frac{1}{m^2}\]
Finally we have
\[\tilde{\Delta}_F^{\nu\rho}(p)=-\frac{i}{p^2+m^2} \left(\eta^{\nu \rho} +\frac{p^\nu p^\rho}{m^2} \right)\]

\smallskip\noindent\textcolor{appmod}{\textbf{Model solution (Gemini-3-Flash-Preview):}}
\begin{center}
$\cdots$ 
(model's reasoning) 
$\cdots$
\end{center}
To determine the momentum-space Feynman propagator $\tilde{\Delta}_F^{\nu\rho}(p)$, we start from the action:
\[S= \int d^4x \, \left( -\frac{1}{4} F_{\mu\nu} F^{\mu \nu}-\frac{1}{2}m^2A_\mu A^\mu\right)\]
Using $F_{\mu \nu} = \partial_\mu A_\nu-\partial_\nu A_\mu$, the Lagrangian density can be rewritten by integrating by parts:
\[\mathcal{L} = \frac{1}{2} A_\nu \left( (\partial^2 - m^2) \eta^{\nu\mu} - \partial^\nu \partial^\mu \right) A_\mu\]
In momentum space, the operator $\tilde{\mathcal{K}}^{\nu\mu}(p)$ corresponding to the quadratic part of the action is obtained by replacing $\partial_\mu \to i p_\mu$:
\[\tilde{\mathcal{K}}^{\nu\mu}(p) = (-p^2 - m^2) \eta^{\nu\mu} + p^\nu p^\mu\]
The propagator $\tilde{\Delta}_{F, \mu\rho}(p)$ is defined as $i$ times the inverse of this operator:
\[\tilde{\mathcal{K}}^{\nu\mu}(p) \tilde{\Delta}_{F, \mu\rho}(p) = i \delta^\nu_\rho\]
Assuming the form $\tilde{\Delta}_{F, \mu\rho}(p) = A \eta_{\mu\rho} + B p_\mu p_\rho$, we substitute it into the equation:
\[[(-p^2 - m^2) \eta^{\nu\mu} + p^\nu p^\mu] (A \eta_{\mu\rho} + B p_\mu p_\rho) = i \delta^\nu_\rho\]
Expanding the terms:
\[-A(p^2 + m^2) \delta^\nu_\rho + (A - B m^2) p^\nu p_\rho = i \delta^\nu_\rho\]
Matching the coefficients for $\delta^\nu_\rho$ and $p^\nu p_\rho$:
1) $-A(p^2 + m^2) = i \implies A = \frac{-i}{p^2 + m^2}$
2) $A - B m^2 = 0 \implies B = \frac{A}{m^2} = \frac{-i}{m^2(p^2 + m^2)}$
This gives the propagator:
\[\tilde{\Delta}_{F, \mu\rho}(p) = \frac{-i}{p^2 + m^2} \left( \eta_{\mu\rho} + \frac{p_\mu p_\rho}{m^2} \right)\]
Raising the indices, we obtain the momentum-space Feynman propagator $\tilde{\Delta}_F^{\nu\rho}(p)$:
\[\tilde{\Delta}_F^{\nu\rho}(p) = \frac{-i}{p^2 + m^2} \left( \eta^{\nu\rho} + \frac{p^\nu p^\rho}{m^2} \right)\]

\[\boxed{\tilde{\Delta}_F^{\nu\rho}(p) = -\frac{i}{p^2 + m^2} \left( \eta^{\nu\rho} + \frac{p^\nu p^\rho}{m^2} \right)}\]
\end{appbox}

\subsection*{QED}

\begin{appbox}
\noindent\textcolor{appq}{\textbf{Question:}}
Compute the differential cross section $\frac{d\sigma}{d\cos\theta}$ for Bhabha scattering at tree level, $e^+ e^- \to e^+ e^-$. Here we ignore the electron mass. Express the result with the fine-structure constant $\alpha$ and the Mandelstam variables $\{s,t,u\}$.

\smallskip\noindent\textcolor{appans}{\textbf{Reference answer:}}
\[
\frac{\pi \alpha^2}{s}\left[\frac{t^2}{s^2}+\frac{s^2}{t^2}+u^2\left(\frac{1}{s}+\frac{1}{t}\right)^2\right]
\]

\smallskip\noindent\textcolor{appsol}{\textbf{Reference solution:}} At tree level the process consists of two diagrams: the $s$-channel annihilation diagram and the $t$-channel exchange diagram (entering with a relative minus sign). The minus sign before the $t$-channel diagram comes from the exchange of two fermion field operators when contracting with in and out states. In the high energy limit, we can ignore the mass of the electrons, so the amplitude for the whole scattering process is
\[i \mathcal{M} = (ie)^2\left[\bar{v}(k_2)\gamma^\mu u(k_1)\frac{-i}{s}\bar{u}(p_1)\gamma_\mu v(p_2)-\bar{u}(p_1)\gamma^\mu u(k_1)\frac{-i}{t}\bar{v}(k_2)\gamma_\mu v(p_2)\right]\]
where the Mandelstam variables $\{s,t,u\}$ are defined as
\[s = (k_1+k_2)^2, \quad t = (p_1-k_1)^2, \quad u = (p_2-k_1)^2\]
In the massless case, $k_1^2=k_2^2=p_1^2=p_2^2=0$, thus we have
\[s=2k_1 \cdot k_2=2p_1 \cdot p_2, \quad t=-2p_1 \cdot k_1=-2p_2 \cdot k_2, \quad u=-2p_2 \cdot k_1=-2p_1 \cdot k_2\]
We want the unpolarized cross section, so we must average over the ingoing spins and sum over the outgoing spins. That is,
\[\begin{aligned}
\frac{1}{4} \sum_{\text{spin}} |\mathcal{M}|^2 &= \frac{e^4}{4s^2} \text{tr}(\slashed{k}_1 \gamma^\mu \slashed{k}_2 \gamma^\nu) \text{tr}(\slashed{p}_2 \gamma_\mu \slashed{p}_1 \gamma_\nu)+\frac{e^4}{4t^2} \text{tr}(\slashed{k}_1 \gamma^\mu \slashed{p}_1 \gamma^\nu) \text{tr}(\slashed{p}_2 \gamma_\mu \slashed{k}_2 \gamma_\nu) \\
&\quad-\frac{e^4}{4st} \left[\text{tr}(\slashed{k}_1 \gamma^\nu \slashed{k}_2 \gamma_\mu \slashed{p}_2 \gamma_\nu \slashed{p}_1 \gamma^\mu)+\text{c.c.} \right] \\
&=\frac{2e^4(u^2+t^2)}{s^2}+\frac{2e^4(u^2+s^2)}{t^2}+\frac{4e^4 u^2}{st} \\
&=2e^4\left[\frac{t^2}{s^2}+\frac{s^2}{t^2}+u^2\left(\frac{1}{s}+\frac{1}{t}\right)^2\right]
\end{aligned}\]
In the center-of-mass frame, we have $k_1^0 =k_2^0 \equiv k^0$ and $\vec{k}_1=-\vec{k}_2$. Thus the total energy $E_{\text{CM}}^2 = (k_1^0+k_2^0)^2 = 4k^2=s$. According to the formula for the cross section of $2\to2$ scattering of massless particles
\[\left(\frac{d\sigma}{d \Omega}\right)_{\text{CM}} = \frac{1}{64\pi^2 E_{\text{CM}}^2}\left(\frac{1}{4} \sum |\mathcal{M}|^2 \right)\]
therefore
\[\left(\frac{d\sigma}{d \Omega}\right)_{\text{CM}} = \frac{\alpha^2}{2s}\left[\frac{t^2}{s^2}+\frac{s^2}{t^2}+u^2\left(\frac{1}{s}+\frac{1}{t}\right)^2\right]\]
where $\alpha = \frac{e^2}{4\pi}$ is the fine structure constant. We integrate this over the angle $\phi$ to get
\[\left( \frac{d\sigma}{d\cos\theta}\right)_{\text{CM}} = \frac{\pi \alpha^2}{s}\left[\frac{t^2}{s^2}+\frac{s^2}{t^2}+u^2\left(\frac{1}{s}+\frac{1}{t}\right)^2\right]\]

\smallskip\noindent\textcolor{appmod}{\textbf{Model solution (Opus-5):}}
\begin{center}
$\cdots$ 
(model's reasoning) 
$\cdots$
\end{center}
\par\smallskip\noindent\textit{Setup}\par

For Bhabha scattering there are two tree diagrams: the \textbf{$t$-channel} (photon exchange between $e^-$ and $e^+$) and the \textbf{$s$-channel} (annihilation into a virtual photon). Summing the two amplitudes, squaring, and averaging over initial spins while summing over final spins gives (massless limit)

\[\overline{|\mathcal{M}|^2}=2e^4\left[\frac{s^2+u^2}{t^2}+\frac{t^2+u^2}{s^2}+\frac{2u^2}{st}\right]\]

where the three terms are $|t\text{-channel}|^2$, $|s\text{-channel}|^2$, and the interference. (Equivalently, this is the M{\o}ller result with $s\leftrightarrow u$ crossed, and can be written compactly as $2e^4\big[u^2(\tfrac1s+\tfrac1t)^2+(s/t)^2+(t/s)^2\big]$.)

\par\smallskip\noindent\textit{Phase space}\par

For $2\to2$ massless elastic scattering in the CM frame,

\[\frac{d\sigma}{d\Omega}=\frac{\overline{|\mathcal{M}|^2}}{64\pi^2 s},
\qquad
\frac{d\sigma}{d\cos\theta}=2\pi\,\frac{d\sigma}{d\Omega}=\frac{\overline{|\mathcal{M}|^2}}{32\pi s}.\]

Using $e^4=(4\pi\alpha)^2=16\pi^2\alpha^2$:

\[\frac{2e^4}{32\pi s}=\frac{e^4}{16\pi s}=\frac{\pi\alpha^2}{s}.\]

\par\smallskip\noindent\textit{Result}\par

\[\boxed{\;\frac{d\sigma}{d\cos\theta}=\frac{\pi\alpha^{2}}{s}\left[\frac{s^{2}+u^{2}}{t^{2}}+\frac{2u^{2}}{st}+\frac{t^{2}+u^{2}}{s^{2}}\right]\;}\]

with $t=-\tfrac{s}{2}(1-\cos\theta)$, $u=-\tfrac{s}{2}(1+\cos\theta)$, so that

\[\frac{d\sigma}{d\cos\theta}=\frac{\pi\alpha^2}{s}\left[\frac{1+\cos^4(\theta/2)}{\sin^4(\theta/2)}-\frac{2\cos^4(\theta/2)}{\sin^2(\theta/2)}+\frac{1+\cos^2\theta}{2}\right].\]

\textbf{Checks:} dropping the $t$-channel and interference terms leaves $\frac{\pi\alpha^2}{s}\frac{t^2+u^2}{s^2}=\frac{\pi\alpha^2}{2s}(1+\cos^2\theta)$, the correct $e^+e^-\to\mu^+\mu^-$ result; the $t\to 0$ ($\theta\to0$) divergence is the expected Rutherford-like forward peak.
\end{appbox}

\subsection*{Renormalization}

\begin{appbox}
\noindent\textcolor{appq}{\textbf{Question:}}
The massless Yukawa Lagrangian is written as
$$
\mathcal{L} = \frac{1}{2}(\partial_\mu \phi)^2-\frac{\lambda}{4!} \phi^4+\sum_{j=1}^N\bar{\psi}_j (i\slashed{\partial})\psi_j-\sum_{j=1}^N ig\bar{\psi}_j \gamma^5 \psi_j \phi
$$
where $\phi$ is a real scalar and $\psi_j$ with $j=1,\cdots,N$ are $N$ Dirac fermions coupled to $\phi$ with common Yukawa coupling $g$. Compute the Callan-Symanzik $\beta$ function for $g$,
$$
\beta_g(\lambda,g),
$$
to leading order in the coupling constants, assuming that $g^2$ and $\lambda$ are of the same order. You only need to consider 1-loop diagrams for the calculation of the counterterms.

\smallskip\noindent\textcolor{appans}{\textbf{Reference answer:}}
\[
\beta_g(\lambda,g)=\frac{(2N+3)g^3}{(4\pi)^2}
\]

\smallskip\noindent\textcolor{appsol}{\textbf{Reference solution:}} Here we consider all the necessary 1-loop diagrams.

The scalar self-energy with a fermion loop (summed over the $N$ flavours):
\[[\text{fermion loop on scalar line}] = -(-ig)^2 \int \frac{d^4 k}{(2\pi)^4} \, \text{tr} \left[ \frac{i}{\slashed{k}} \gamma^5 \frac{i}{\slashed{k}-\slashed{p}}\gamma^5\right] \sim \frac{4iNg^2 p^2}{(4\pi)^2} \frac{1}{\epsilon}\]
The fermion self-energy with a scalar exchange:
\[[\text{scalar loop on fermion line}] = g^2 \int \frac{d^4 k}{(2\pi)^4} \, \gamma^5 \frac{i}{\slashed{k}}\gamma^5 \frac{i}{(k-p)^2} \sim \frac{ig^2 \slashed{p}}{(4\pi)^2}\]
The vertex correction:
\[[\text{Yukawa vertex correction}] = g^3 \int \frac{d^4 k}{(2\pi)^4} \, \gamma^5 \frac{i}{\slashed{k}}\gamma^5 \frac{i}{\slashed{k}}\gamma^5 \frac{i}{k^2} \sim -i\frac{g^3 \gamma^5}{(4\pi)^2} \frac{1}{\epsilon}\]
We can find the counterterms as follows (note that there are two permutations for the Yukawa loop):
\[\delta_\phi = -\frac{2Ng^2 }{(4\pi)^2}\left( \frac{2}{\epsilon} -\ln \mu^2 \right)\]
\[\delta_\psi = -\frac{g^2}{2(4\pi)^2} \left( \frac{2}{\epsilon} -\ln \mu^2 \right)\]
\[\delta_g = \frac{g^3}{(4\pi)^2} \left( \frac{2}{\epsilon} -\ln \mu^2 \right)\]
where $\mu$ is the renormalization scale. The beta function for $g$ to lowest order is given by
\[\beta_g(\lambda,g) = \mu \frac{\partial}{\partial\mu} \left(-\delta_g+\frac{1}{2}g_0\delta_\phi+g_0 \delta_\psi\right)=\frac{(2N+3)g^3}{(4\pi)^2}\]

\smallskip\noindent\textcolor{appmod}{\textbf{Model solution (Opus-5):}}
\begin{center}
$\cdots$ 
(model's reasoning) 
$\cdots$
\end{center}
\par\smallskip\noindent\textit{Setup}\par

With $\mathcal{L}_{\rm int}=-ig\bar\psi\gamma^5\psi\phi$ the vertex Feynman rule is $g\gamma^5$. Writing the counterterms as
\[\tfrac12\delta_\phi(\partial\phi)^2+\delta_\psi\bar\psi i\slashed\partial\psi-i\delta_g\bar\psi\gamma^5\psi\phi ,\]
the bare coupling is $g_0=(g+\delta_g)Z_\psi^{-1}Z_\phi^{-1/2}$, so that
\[\beta_g=-M\frac{d}{dM}\Big[\delta_g-g\,\delta_\psi-\tfrac{g}{2}\,\delta_\phi\Big].\]
(In dim. reg. $2/\epsilon \leftrightarrow \log\Lambda^2/M^2$, so $M\frac{d}{dM}\frac{2}{\epsilon}\to-2$.)

\par\smallskip\noindent\textit{One-loop counterterms}\par

\textbf{Scalar self-energy} (fermion loop, factor $N$, using $\mathrm{tr}[\gamma^5\slashed k\gamma^5(\slashed k+\slashed p)]=-4k\cdot(k+p)$):
\[-i\Sigma=-4Ng^2\!\int\!\frac{d^dk}{(2\pi)^d}\frac{k\cdot(k+p)}{k^2(k+p)^2}\Big|_{\rm div}=\frac{i\,4Ng^2p^2}{(4\pi)^2}\frac{1}{\epsilon}\cdot 1
\;\Rightarrow\;\delta_\phi=-\frac{2Ng^2}{(4\pi)^2}\frac{2}{\epsilon}\]

\textbf{Fermion self-energy:}
\[-i\Sigma_\psi=g^2\!\int\!\frac{d^dk}{(2\pi)^d}\frac{\slashed k}{k^2(p-k)^2}\Big|_{\rm div}=\frac{i\,g^2\slashed p}{2(4\pi)^2}\frac{2}{\epsilon}
\;\Rightarrow\;\delta_\psi=-\frac{g^2}{2(4\pi)^2}\frac{2}{\epsilon}\]

\textbf{Vertex correction} (using $\gamma^5\slashed k\gamma^5\slashed k\gamma^5=-k^2\gamma^5$):
\[V=i^3g^3\!\int\!\frac{d^dk}{(2\pi)^d}\frac{-k^2\gamma^5}{(k^2)^3}=-\frac{g^3\gamma^5}{(4\pi)^2}\frac{2}{\epsilon}
\;\Rightarrow\;\delta_g=+\frac{g^3}{(4\pi)^2}\frac{2}{\epsilon}\]

There is \textbf{no} $O(\lambda g)$ contribution: attaching the $\phi^4$ vertex to the fermion line requires at least two loops (the tadpole contraction vanishes).

\par\smallskip\noindent\textit{Beta function}\par

\[M\frac{d\delta_g}{dM}=-\frac{2g^3}{(4\pi)^2},\qquad 
M\frac{d\delta_\psi}{dM}=\frac{g^2}{(4\pi)^2},\qquad 
M\frac{d\delta_\phi}{dM}=\frac{4Ng^2}{(4\pi)^2}\]

\[\beta_g=\frac{2g^3}{(4\pi)^2}+\frac{g^3}{(4\pi)^2}+\frac{2Ng^3}{(4\pi)^2}\]

\[\boxed{\;\beta_g(\lambda,g)=\frac{(2N+3)\,g^{3}}{16\pi^{2}}+O(g^5,\lambda g^3)\;}\]

(For $N=1$ this reproduces the familiar $\beta_g=5g^3/16\pi^2$; the coupling is not asymptotically free.)
\end{appbox}

\subsection*{Non-Abelian}

\begin{appbox}
\noindent\textcolor{appq}{\textbf{Question:}}
Consider a non-Abelian Yang-Mills theory with gauge field $A_{\mu}^{a}$ and field strength
$$
F_{\mu\nu}^a
=
\partial_\mu A_\nu^a-\partial_\nu A_\mu^a
+g f^{abc}A_\mu^b A_\nu^c .
$$
Choose the convention $\epsilon^{0123}=1$. The $\theta$-term is given by $\mathcal{L}_\theta=\epsilon^{\mu\nu\lambda\rho}F_{\mu\nu}^a F_{\lambda\rho}^a$. Show that $\mathcal{L}_\theta$ can be written locally as a total derivative $\mathcal{L}_\theta=\partial_{\mu}K^{\mu}$, and determine the standard current $K^{\mu}$ explicitly in terms of the gauge field $A^{a}_{\mu}$ and field strength $F^{a}_{\mu\nu}$.

\smallskip\noindent\textcolor{appans}{\textbf{Reference answer:}}
\[
2\epsilon^{\mu\nu\lambda\rho}\left(A_\nu^aF_{\lambda\rho}^a-\frac{g}{3}f^{abc}A_\nu^aA_\lambda^bA_\rho^c\right)
\]

\smallskip\noindent\textcolor{appsol}{\textbf{Reference solution:}} We want to write $\epsilon^{\mu\nu\lambda\rho}F_{\mu\nu}^aF_{\lambda\rho}^a$ as a total derivative. A natural first guess is
\[K_0^\mu
=
2\epsilon^{\mu\nu\lambda\rho}A_\nu^aF_{\lambda\rho}^a .\]
Taking the divergence gives
\[\partial_\mu K_0^\mu
=
2\epsilon^{\mu\nu\lambda\rho}
(\partial_\mu A_\nu^a)F_{\lambda\rho}^a
+
2\epsilon^{\mu\nu\lambda\rho}
A_\nu^a\partial_\mu F_{\lambda\rho}^a .\]
After using
\[F_{\mu\nu}^a
=
\partial_\mu A_\nu^a-\partial_\nu A_\mu^a
+
g f^{abc}A_\mu^bA_\nu^c\]
and relabelling dummy indices, this becomes
\[\partial_\mu K_0^\mu
=
\epsilon^{\mu\nu\lambda\rho}
F_{\mu\nu}^aF_{\lambda\rho}^a
+
2g\epsilon^{\mu\nu\lambda\rho}
f^{abc}
A_\mu^aA_\nu^b\partial_\lambda A_\rho^c .\]
Thus the naive current gives the desired term plus an extra contribution. This extra contribution is itself a total derivative:
\[2g\epsilon^{\mu\nu\lambda\rho}
f^{abc}
A_\mu^aA_\nu^b\partial_\lambda A_\rho^c
=
\frac{2g}{3}\partial_\mu
\left(
\epsilon^{\mu\nu\lambda\rho}
f^{abc}A_\nu^aA_\lambda^bA_\rho^c
\right).\]
Therefore the correct current is obtained by subtracting this extra piece
\[K^\mu
=
2\epsilon^{\mu\nu\lambda\rho}
A_\nu^aF_{\lambda\rho}^a
-
\frac{2g}{3}
\epsilon^{\mu\nu\lambda\rho}
f^{abc}A_\nu^aA_\lambda^bA_\rho^c.\]
Equivalently,
\[K^\mu
=
2\epsilon^{\mu\nu\lambda\rho}
\left(
A_\nu^aF_{\lambda\rho}^a
-
\frac{g}{3}f^{abc}A_\nu^aA_\lambda^bA_\rho^c
\right).\]
With this choice, $\mathcal{L}_\theta$ is a total derivative.

\smallskip\noindent\textcolor{appmod}{\textbf{Model solution (Gemini-3.8-Flash):}}
\begin{center}
$\cdots$ 
(model's reasoning) 
$\cdots$
\end{center}
To show that $\mathcal{L}_\theta = \epsilon^{\mu\nu\lambda\rho} F_{\mu\nu}^a F_{\lambda\rho}^a$ can be written locally as a total derivative $\mathcal{L}_\theta = \partial_\mu K^\mu$, we can work directly with the field strength components or differential forms.

\par\smallskip\noindent\textit{1. Derivation}\par

The non-Abelian field strength is given by:
\[F_{\mu\nu}^a = \partial_\mu A_\nu^a - \partial_\nu A_\mu^a + g f^{abc} A_\mu^b A_\nu^c\]

Substituting this into $\mathcal{L}_\theta$, we expand the product:
\[\mathcal{L}_\theta = \epsilon^{\mu\nu\lambda\rho} F_{\mu\nu}^a F_{\lambda\rho}^a = \epsilon^{\mu\nu\lambda\rho} \left(2\partial_\mu A_\nu^a + g f^{abc} A_\mu^b A_\nu^c\right) \left(2\partial_\lambda A_\rho^a + g f^{ade} A_\lambda^d A_\rho^e\right)\]

Expanding the terms:
1. \textbf{Kinetic term:}
   \[4 \epsilon^{\mu\nu\lambda\rho} (\partial_\mu A_\nu^a)(\partial_\lambda A_\rho^a) = 4 \partial_\mu \left(\epsilon^{\mu\nu\lambda\rho} A_\nu^a \partial_\lambda A_\rho^a\right)\]
   (since $\epsilon^{\mu\nu\lambda\rho} \partial_\mu \partial_\lambda A_\rho^a = 0$ by the commutativity of partial derivatives).

2. \textbf{Cross terms:}
   \[4 g f^{abc} \epsilon^{\mu\nu\lambda\rho} (\partial_\mu A_\nu^a) A_\lambda^b A_\rho^c\]
   Notice that:
   \[\partial_\mu \left(\epsilon^{\mu\nu\lambda\rho} f^{abc} A_\nu^a A_\lambda^b A_\rho^c\right) = 3 \epsilon^{\mu\nu\lambda\rho} f^{abc} (\partial_\mu A_\nu^a) A_\lambda^b A_\rho^c\]
   due to the cyclic symmetry of $f^{abc}$ and the antisymmetry of $\epsilon^{\mu\nu\lambda\rho}$. Thus,
   \[4 g f^{abc} \epsilon^{\mu\nu\lambda\rho} (\partial_\mu A_\nu^a) A_\lambda^b A_\rho^c = \partial_\mu \left( \frac{4}{3} g f^{abc} \epsilon^{\mu\nu\lambda\rho} A_\nu^a A_\lambda^b A_\rho^c \right)\]

3. \textbf{Quartic term:}
   \[g^2 \epsilon^{\mu\nu\lambda\rho} f^{abc} f^{ade} A_\mu^b A_\nu^c A_\lambda^d A_\rho^e = 0\]
   This vanishes identically due to the Jacobi identity: $f^{a[bc} f^{de]a} = 0$.

Combining these results, we find:
\[\mathcal{L}_\theta = \partial_\mu \left[ 4 \epsilon^{\mu\nu\lambda\rho} \left( A_\nu^a \partial_\lambda A_\rho^a + \frac{1}{3} g f^{abc} A_\nu^a A_\lambda^b A_\rho^c \right) \right]\]

\par\smallskip\noindent\textit{2. Rewriting in terms of $A_\mu^a$ and $F_{\mu\nu}^a$}\par

Using the relation $\partial_\lambda A_\rho^a = \frac{1}{2} \left( F_{\lambda\rho}^a - g f^{abc} A_\lambda^b A_\rho^c \right)$ (valid when contracted with $\epsilon^{\mu\nu\lambda\rho}$), we substitute it into the bracketed current:
\[A_\nu^a \partial_\lambda A_\rho^a + \frac{1}{3} g f^{abc} A_\nu^a A_\lambda^b A_\rho^c = \frac{1}{2} A_\nu^a F_{\lambda\rho}^a - \frac{1}{6} g f^{abc} A_\nu^a A_\lambda^b A_\rho^c\]

Multiplying by $4 \epsilon^{\mu\nu\lambda\rho}$, we obtain:
\[K^\mu = 2 \epsilon^{\mu\nu\lambda\rho} \left( A_\nu^a F_{\lambda\rho}^a - \frac{1}{3} g f^{abc} A_\nu^a A_\lambda^b A_\rho^c \right)\]

This is the standard topological Chern-Simons current.

\[\boxed{K^{\mu} = 2 \epsilon^{\mu\nu\lambda\rho} \left( A_\nu^a F_{\lambda\rho}^a - \frac{1}{3} g f^{abc} A_\nu^a A_\lambda^b A_\rho^c \right)}\]
\end{appbox}

\subsection*{Electroweak}

\begin{appbox}
\noindent\textcolor{appq}{\textbf{Question:}}
The Higgs Yukawa coupling to fermions is given by
$$
\mathcal{L}_Y =\frac{g}{2}\frac{m_f}{M_W}h\bar{f}f .
$$
Find the decay rate $\Gamma$ for the Higgs into a fermion-antifermion pair at tree level. Take the mass of $h$ to be $m_h$. Account for the situation $f=\text{quark}$; the number of colors can be written as $N_c$.

\smallskip\noindent\textcolor{appans}{\textbf{Reference answer:}}
\[
\Gamma=m_h \frac{N_c g^2m_f^2}{32\pi M_W^2}\left(1-\frac{4m_f^2}{m_h^2}\right)^{\frac{3}{2}}
\]

\smallskip\noindent\textcolor{appsol}{\textbf{Reference solution:}} To keep the expression short we introduce $y=\frac{g}{2}\frac{m_f}{M_W}$; the tree-level amplitude reads
\[\mathcal{M} = y\bar{u}_{s_1}(\vec{p}_1)v_{s_2}(\vec{p}_2)\]
with $\{\vec{p}_i,s_i\}$ the momenta and spins of the outgoing fermion and anti-fermion. Then
\[\begin{aligned}
\sum_{\text{spin}} |\mathcal{M}|^2 &=y^2 \sum_{\text{spin}} \bar{u}_{s_1}(\vec{p}_1)\,v_{s_2}(\vec{p}_2)\,\bar v_{s_2}(\vec{p}_2)\,u_{s_1}(\vec{p}_1) \\
&=y^2\text{Tr}[(\slashed{p}_2-m_f)(\slashed{p}_1+m_f)] =y^2\text{Tr}[\slashed{p}_2 \slashed{p}_1-m_f^2] =4y^2(p_1 \cdot p_2-m_f^2)
\end{aligned}\]
In the center-of-mass frame,
\[k^\mu=(m_h,\vec{0}), \quad p^\mu_1 = \left(\frac{m_h}{2},\vec{p}\right), \quad p^\mu_2 = \left(\frac{m_h}{2},-\vec{p}\right)\]
Since from 4-momentum conservation $m_h=2E_f$ and $E_f^2 = \vec{p}^2+m_f^2$, then
\[\sum_{\text{spin}} |\mathcal{M}|^2=4y^2\left(\frac{m_h^2}{2}-2m_f^2\right)\]
Having computed the amplitude, the total decay rate is
\[\Gamma = \frac{N_c}{8\pi m_h^2}|\vec{p}|\,|\mathcal{M}|^2\]
Since $|\vec{p}|=\frac{m_h}{2}\left(1-\frac{4m_f^2}{m_h^2}\right)^{\frac{1}{2}}$, therefore
\[\Gamma=m_h \frac{N_c g^2m_f^2}{32\pi M_W^2}\left(1-\frac{4m_f^2}{m_h^2}\right)^{\frac{3}{2}}\]

\smallskip\noindent\textcolor{appmod}{\textbf{Model solution (Gemini-3.8-Flash):}}
\begin{center}
$\cdots$ 
(model's reasoning) 
$\cdots$
\end{center}
To find the tree-level decay rate $\Gamma(h \to f\bar{f})$, we follow these steps:

\par\smallskip\noindent\textit{1. Matrix Element}\par
From the Yukawa interaction term in the Lagrangian,
\[\mathcal{L}_Y = \frac{g}{2}\frac{m_f}{M_W}h\bar{f}f,\]
the invariant amplitude for the decay $h(p) \to f(p_1) + \bar{f}(p_2)$ is:
\[\mathcal{M} = \frac{g m_f}{2 M_W} \bar{u}(p_1) v(p_2).\]

\par\smallskip\noindent\textit{2. Spin Sum and Color Factor}\par
Summing over the final fermion and antifermion spins and colors (with $N_c$ colors, where $N_c = 3$ for quarks and $N_c = 1$ for leptons):
\[\sum_{\text{spins, colors}} |\mathcal{M}|^2 = N_c \left(\frac{g m_f}{2 M_W}\right)^2 \operatorname{Tr}\left[ (\not{p}_1 + m_f)(\not{p}_2 - m_f) \right].\]
Using the trace identities:
\[\operatorname{Tr}\left[ (\not{p}_1 + m_f)(\not{p}_2 - m_f) \right] = \operatorname{Tr}(\not{p}_1 \not{p}_2) - 4 m_f^2 = 4(p_1 \cdot p_2 - m_f^2).\]
From four-momentum conservation in the rest frame of the Higgs, $p = p_1 + p_2$, so:
\[m_h^2 = p^2 = (p_1 + p_2)^2 = 2m_f^2 + 2p_1 \cdot p_2 \implies p_1 \cdot p_2 = \frac{m_h^2}{2} - m_f^2.\]
Substituting this back gives:
\[\operatorname{Tr}\left[ (\not{p}_1 + m_f)(\not{p}_2 - m_f) \right] = 4\left(\frac{m_h^2}{2} - 2m_f^2\right) = 2m_h^2 \left(1 - \frac{4m_f^2}{m_h^2}\right).\]
Therefore:
\[\sum_{\text{spins, colors}} |\mathcal{M}|^2 = 2 N_c \left(\frac{g m_f}{2 M_W}\right)^2 m_h^2 \left(1 - \frac{4m_f^2}{m_h^2}\right).\]

\par\smallskip\noindent\textit{3. Phase Space and Decay Rate}\par
The two-body decay width in the Higgs rest frame is:
\[\Gamma = \frac{|\vec{p}_1|}{8\pi m_h^2} \sum |\mathcal{M}|^2,\]
where the magnitude of the final-state fermion momentum is:
\[|\vec{p}_1| = \frac{m_h}{2}\sqrt{1 - \frac{4m_f^2}{m_h^2}}.\]

Combining the factors:
\[\Gamma = \frac{1}{8\pi m_h^2} \left[ \frac{m_h}{2}\sqrt{1 - \frac{4m_f^2}{m_h^2}} \right] \left[ 2 N_c \frac{g^2 m_f^2}{4 M_W^2} m_h^2 \left(1 - \frac{4m_f^2}{m_h^2}\right) \right].\]

Simplifying the expression:
\[\Gamma = \frac{N_c g^2 m_f^2 m_h}{32\pi M_W^2} \left(1 - \frac{4m_f^2}{m_h^2}\right)^{3/2}.\]

\textit{(Note: Using the relation $\frac{G_F}{\sqrt{2}} = \frac{g^2}{8 M_W^2}$, this can also be equivalently written as $\Gamma = \frac{N_c G_F m_f^2 m_h}{4\sqrt{2}\pi} \left(1 - \frac{4m_f^2}{m_h^2}\right)^{3/2}$.)}

\[\boxed{\Gamma = \frac{N_c g^2 m_f^2 m_h}{32 \pi M_W^2} \left(1 - \frac{4m_f^2}{m_h^2}\right)^{3/2}}\]
\end{appbox}

\subsection*{SUSY}

\begin{appbox}
\noindent\textcolor{appq}{\textbf{Question:}}
In pure $SU(2)$ Seiberg-Witten theory, the Coulomb branch has a $\mathbb{Z}_{2}$ symmetry under $u\mapsto -u$. The moduli space has three singularities located at
$$
u=\infty, \qquad u=u_{+}, \qquad u=u_{-}=-u_{+}
$$
At $u=u_{+}$, a monopole with electromagnetic charge $(n_{m},n_{e})=(1,0)$ becomes massless, while at $u=u_{-}$, a dyon with charge $(n_{m},n_{e})=(1,-1)$ becomes massless. The central charge is $Z=n_{m}a_{D}+n_{e}a$. Using the period basis $(a_{D},a)^{T}$ and counterclockwise monodromy loops, compute the ordered pair of monodromy matrices $(M_{u_{+}},M_{u_{-}})$ associated with the two finite singularities $u_{+}$ and $u_{-}$.

\smallskip\noindent\textcolor{appans}{\textbf{Reference answer:}}
\[
(M_{u_{+}},M_{u_{-}})=\left(\begin{pmatrix} 1 & 0\\ -2 & 1 \end{pmatrix},\,\begin{pmatrix} -1 & 2\\ -2 & 3 \end{pmatrix} \right)
\]

\smallskip\noindent\textcolor{appsol}{\textbf{Reference solution:}} If $u=\infty$ were the only singularity, then the monodromy around infinity could be continuously deformed to a trivial contour in the finite $u$-plane. This would imply that the periods $(a_{D},a)^{T}$ have no nontrivial monodromy, so $a$ would be a good global coordinate on the whole moduli space. However, this is not true: the Seiberg-Witten periods have nontrivial monodromy, and the quantum moduli space must contain finite singularities where extra BPS states become massless. At the monopole singularity $u=u_{+}$, a monopole with charge $(n_{m},n_{e})=(1,0)$ becomes massless. Since the BPS central charge is $Z=n_{m}a_{D}+n_{e}a$, the massless condition implies $a_{D}=0$. Near the singularity $u=u_{+}$, the dual coupling behaves as
\[\tau_D=\frac{da}{da_D}
\sim
\frac{i}{\pi}\log a_D.\]
Integrating, we obtain
\[a
\sim
a_{+}
+\frac{i}{\pi}a_D\log a_D,\]
where $a_{+}=a(u_{+})$ is the finite value of $a$ at the monopole point. For $u-u_{+}\mapsto e^{2\pi i}(u-u_{+})$, we obtain
\[a_{D}\mapsto a_{D}, \qquad a\mapsto a-2a_{D}\]
So the monodromy matrix at $u=u_{+}$ is
\[M_{u_{+}}
=
\begin{pmatrix}
1 & 0\\
-2 & 1
\end{pmatrix}\]
From the large-$u$ asymptotic behavior, the monodromy at infinity is
\[M_\infty=
\begin{pmatrix}
-1 & 2\\
0 & -1
\end{pmatrix}.\]
For the chosen counterclockwise loops, the monodromies satisfy
\[M_\infty=M_{u_+}M_{u_-}.\]
Therefore,
\[M_{u_-}
=
M_{u_+}^{-1}M_\infty
=
\begin{pmatrix}
-1 & 2\\
-2 & 3
\end{pmatrix}.\]

\smallskip\noindent\textcolor{appmod}{\textbf{Model solution (Gemini-3.6-Flash):}}
\begin{center}
$\cdots$ 
(model's reasoning) 
$\cdots$
\end{center}
To find the monodromy matrices $M_{u_+}$ and $M_{u_-}$ in pure $SU(2)$ Seiberg-Witten theory, we use the Picard-Lefschetz formula for the monodromy acting on the period basis $v = \begin{pmatrix} a_D \\ a \end{pmatrix}$.

When circling counterclockwise around a singularity where a hypermultiplet with charge $q = \begin{pmatrix} n_m \\ n_e \end{pmatrix}$ becomes massless, the monodromy matrix $M_q$ acts on $v$ as:
\[M_q v = v - 2 (n_m a_D + n_e a) \begin{pmatrix} -n_e \\ n_m \end{pmatrix}\]
In matrix form, $M_q$ is given by:
\[M_q = \begin{pmatrix} 1 + 2 n_e n_m & 2 n_e^2 \\ -2 n_m^2 & 1 - 2 n_e n_m \end{pmatrix}\]

1. \textbf{For $u = u_+$:} 
   The massless state is a monopole with charge $(n_m, n_e) = (1, 0)$. Substituting these values into the matrix formula gives:
   \[   M_{u_+} = \begin{pmatrix} 1 + 2(0)(1) & 2(0)^2 \\ -2(1)^2 & 1 - 2(0)(1) \end{pmatrix} = \begin{pmatrix} 1 & 0 \\ -2 & 1 \end{pmatrix}
   \]

2. \textbf{For $u = u_-$:} 
   The massless state is a dyon with charge $(n_m, n_e) = (1, -1)$. Substituting these values gives:
   \[   M_{u_-} = \begin{pmatrix} 1 + 2(-1)(1) & 2(-1)^2 \\ -2(1)^2 & 1 - 2(-1)(1) \end{pmatrix} = \begin{pmatrix} -1 & 2 \\ -2 & 3 \end{pmatrix}
   \]

As a consistency check, the product $M_{u_+} M_{u_-} = \begin{pmatrix} -1 & 2 \\ 0 & -1 \end{pmatrix} = M_{\infty}$, which matches the expected monodromy at infinity due to asymptotic freedom.

\[\boxed{\left( \begin{pmatrix} 1 & 0 \\ -2 & 1 \end{pmatrix}, \begin{pmatrix} -1 & 2 \\ -2 & 3 \end{pmatrix} \right)}\]
\end{appbox}

\subsection*{CFT}

\begin{appbox}
\noindent\textcolor{appq}{\textbf{Question:}}
Let $\phi$ and $\mathcal{O}$ be scalar primary operators in a $d$-dimensional CFT with scaling dimensions $\Delta_\phi$ and $\Delta$, respectively. Assume that the two-point function is normalized as
$$
\langle \mathcal O(x)\mathcal O(y)\rangle
=
\frac{1}{|x-y|^{2\Delta}}.
$$
Write the $\phi\times\phi$ OPE as
$$
\phi(x)\phi(0)=f_{\phi\phi\mathcal{O}}C(x,\partial)\mathcal{O}(0)+\cdots,
$$
where the differential operator $C(x,\partial)$ has the expansion
$$
C(x,\partial)
=
|x|^{\Delta-2\Delta_\phi}
\left(
1+\alpha\,x\cdot \partial
+\beta x^\mu x^\nu \partial_\mu\partial_\nu
+\gamma x^2\partial^2+\cdots
\right).
$$
Assume that $\Delta\neq\frac{d-2}{2}$. Determine the coefficient tuple $(\alpha,\beta,\gamma)$.

\smallskip\noindent\textcolor{appans}{\textbf{Reference answer:}}
\[
(\alpha,\beta,\gamma)=\left(\frac12,\frac{\Delta+2}{8(\Delta+1)},-\frac{\Delta}{16\left(\Delta-\frac{d-2}{2}\right)(\Delta+1)}\right)
\]

\smallskip\noindent\textcolor{appsol}{\textbf{Reference solution:}} By conformal invariance, the three-point function of scalar operators is fixed to be
\[\langle \phi(x)\phi(0)\mathcal O(z)\rangle = \frac{f_{\phi\phi\mathcal O}} {|x|^{2\Delta_\phi-\Delta}|z|^\Delta |z-x|^\Delta} = \frac{f_{\phi\phi\mathcal O} |x|^{\Delta-2\Delta_\phi}}{|z|^\Delta |z-x|^\Delta}.\]
On the other hand, using the OPE, we obtain
\[\langle \phi(x)\phi(0)\mathcal O(z)\rangle = f_{\phi\phi\mathcal O}\, C(x,\partial_{y}) \langle \mathcal O(y)\mathcal O(z)\rangle\big|_{y=0}\]
We choose the normalization
\[\langle \mathcal O(y)\mathcal O(z)\rangle = \frac{1}{|y-z|^{2\Delta}}.\]
Therefore, $C(x,\partial_{y})$ must satisfy
\[\left.
\frac{C(x,\partial_{y})}{|x|^{\Delta-2\Delta_\phi}}\frac{1}{|z-y|^{2\Delta}}\right|_{y=0}= \frac{1}{|z|^\Delta |z-x|^\Delta}.\]
Now we expand the right-hand side
\[\frac{1}{|z|^\Delta |z-x|^\Delta} = \frac{1}{|z|^{2\Delta}} \left[ 1 +\Delta\frac{x\cdot z}{z^2} -\frac{\Delta}{2}\frac{x^2}{z^2} +\frac{\Delta(\Delta+2)}{2} \frac{(x\cdot z)^2}{z^4} +\cdots \right].\]
The left-hand side gives
\[\frac{1}{|z|^{2\Delta}}
\left[
1
+2\alpha\Delta \frac{x\cdot z}{z^2}
+4\beta\Delta(\Delta+1)\frac{(x\cdot z)^2}{z^4}
+\left(
-2\beta\Delta
+2\gamma\Delta(2\Delta+2-d)
\right)\frac{x^2}{z^2}
+\cdots
\right].\]
Comparing coefficients gives
\[\alpha=\frac12,
\qquad
\beta=\frac{\Delta+2}{8(\Delta+1)},
\qquad
\gamma=
-\frac{\Delta}
{16\left(\Delta-\frac{d-2}{2}\right)(\Delta+1)}.\]

\smallskip\noindent\textcolor{appmod}{\textbf{Model solution (Gemini-3.8-Flash):}}
\begin{center}
$\cdots$ 
(model's reasoning) 
$\cdots$
\end{center}
To determine the coefficients $\alpha$, $\beta$, and $\gamma$, we consider the three-point correlation function of scalar primary operators:
\[\langle \phi(x)\phi(0)\mathcal{O}(x_3) \rangle = \frac{f_{\phi\phi\mathcal{O}}}{|x|^{2\Delta_\phi - \Delta} |x_3|^{\Delta} |x - x_3|^{\Delta}}.\]
Applying the $\phi \times \phi$ OPE:
\[\phi(x)\phi(0) = f_{\phi\phi\mathcal{O}} C(x,\partial) \mathcal{O}(0) + \cdots,\]
we have:
\[\langle \phi(x)\phi(0)\mathcal{O}(x_3) \rangle = f_{\phi\phi\mathcal{O}} C(x,\partial_y) \langle \mathcal{O}(y)\mathcal{O}(x_3) \rangle \Big|_{y=0} = f_{\phi\phi\mathcal{O}} C(x,\partial_y) \frac{1}{|y-x_3|^{2\Delta}}\Big|_{y=0}.\]
Matching this with the exact three-point function requires:
\[\left( 1 + \alpha\, x\cdot \partial_y + \beta\, x^\mu x^\nu \partial_\mu \partial_\nu + \gamma\, x^2 \partial^2 + \cdots \right) \frac{1}{|y-x_3|^{2\Delta}}\Bigg|_{y=0} = \frac{1}{|x_3|^\Delta |x_3 - x|^\Delta}.\]

\par\smallskip\noindent\textit{1. Expansion of the RHS}\par
Let $X = x_3$. Expanding $|X-x|^{-\Delta}$ in powers of $x$:
\[|X-x|^{-\Delta} = |X|^{-\Delta} \left(1 - 2\frac{x\cdot X}{X^2} + \frac{x^2}{X^2}\right)^{-\Delta/2}
= |X|^{-\Delta} \left[ 1 + \Delta \frac{x\cdot X}{X^2} - \frac{\Delta}{2}\frac{x^2}{X^2} + \frac{\Delta(\Delta+2)}{2}\frac{(x\cdot X)^2}{X^4} + \mathcal{O}(x^3) \right].\]
Thus, the RHS becomes:
\[\frac{1}{|X|^{2\Delta}} \left[ 1 + \Delta \frac{x\cdot X}{X^2} - \frac{\Delta}{2}\frac{x^2}{X^2} + \frac{\Delta(\Delta+2)}{2}\frac{(x\cdot X)^2}{X^4} + \mathcal{O}(x^3) \right].\]

\par\smallskip\noindent\textit{2. Expansion of the LHS}\par
Let $f(y) = |y-X|^{-2\Delta}$. Computing the derivatives evaluated at $y=0$:
\begin{itemize}\setlength\itemsep{0pt}
\item \textbf{First derivative:}
  \[  \partial_\mu f(0) = 2\Delta \frac{X_\mu}{X^2} f(0).
  \]
\item \textbf{Second derivative:}
  \[  \partial_\mu \partial_\nu f(0) = f(0) \left[ -2\Delta \frac{\delta_{\mu\nu}}{X^2} + 4\Delta(\Delta+1) \frac{X_\mu X_\nu}{X^4} \right].
  \]
\item \textbf{Laplacian:}
  \[  \partial^2 f(0) = \delta^{\mu\nu} \partial_\mu \partial_\nu f(0) = f(0) \frac{2\Delta(2\Delta - d + 2)}{X^2}.
  \]
\end{itemize}

Acting on $f(0)$ with the differential operators:
\begin{itemize}\setlength\itemsep{0pt}
\item Linear order in $x$:
  \[  \alpha\, x^\mu \partial_\mu f(0) = 2\alpha\Delta \frac{x\cdot X}{X^2} f(0).
  \]
\item Quadratic order in $x$:
  \[  \left( \beta x^\mu x^\nu \partial_\mu\partial_\nu + \gamma x^2 \partial^2 \right) f(0) = f(0) \left[ 4\Delta(\Delta+1)\beta \frac{(x\cdot X)^2}{X^4} + 2\Delta \left( -\beta + (2\Delta - d + 2)\gamma \right) \frac{x^2}{X^2} \right].
  \]
\end{itemize}

\par\smallskip\noindent\textit{3. Matching Coefficients}\par
\begin{itemize}\setlength\itemsep{0pt}
\item Matching the $x\cdot X$ term:
  \[  2\alpha\Delta = \Delta \implies \alpha = \frac{1}{2}.
  \]
\item Matching the $(x\cdot X)^2$ term:
  \[  4\Delta(\Delta+1)\beta = \frac{\Delta(\Delta+2)}{2} \implies \beta = \frac{\Delta+2}{8(\Delta+1)}.
  \]
\item Matching the $x^2$ term:
  \[  2\Delta \left( -\beta + (2\Delta - d + 2)\gamma \right) = -\frac{\Delta}{2} \implies -\beta + (2\Delta - d + 2)\gamma = -\frac{1}{4}.
  \]
  Substituting $\beta$:
  \[  (2\Delta - d + 2)\gamma = \frac{\Delta+2}{8(\Delta+1)} - \frac{1}{4} = -\frac{\Delta}{8(\Delta+1)}.
  \]
  Since $\Delta \neq \frac{d-2}{2}$, we obtain:
  \[  \gamma = -\frac{\Delta}{8(\Delta+1)(2\Delta - d + 2)}.
  \]
\end{itemize}

\[\boxed{\left(\frac{1}{2}, \frac{\Delta+2}{8(\Delta+1)}, -\frac{\Delta}{8(\Delta+1)(2\Delta-d+2)}\right)}\]
\end{appbox}